\documentclass[twocolumn,secnumarabic,amssymb, nobibnotes, aps, prl, superscriptaddress, nobalancelastpage,longbibliography,nofootinbib]{revtex4-1}

 \usepackage{graphicx}
\usepackage{bm}
\usepackage{amsmath} \usepackage{braket} \usepackage{epsfig}
\usepackage{tensor}
\usepackage[version=4]{mhchem}
\usepackage{titlesec}
\usepackage{ifthen}
\usepackage{sidecap}
\usepackage{listings}
\usepackage[para,online,flushleft]{threeparttablex}

\usepackage{xr-hyper}
\usepackage{hyperref}
\hypersetup{breaklinks=true,colorlinks=true,linkcolor=blue,citecolor=blue,filecolor=magenta,urlcolor=cyan}

\usepackage[all]{hypcap}

\usepackage{ulem}

\usepackage{xcolor}
\definecolor{pastelgray}{rgb}{0.81, 0.81, 0.77}
\definecolor{beaublue}{rgb}{0.9, 0.9, 0.93}

\makeatletter
\def\@bibdataout@aps{%
\immediate\write\@bibdataout{%
@CONTROL{%
apsrev41Control%
\longbibliography@sw{%
    ,author="08",editor="1",pages="1",title="0",year="1"%
    }{%
    ,author="08",editor="1",pages="1",title="",year="1"%
    }%
  }%
}%
\if@filesw \immediate \write \@auxout {\string \citation {apsrev41Control}}\fi
}
\makeatother

\newcommand{\Apv}{A_{\rm PV}}
\newcommand{\aD}{\alpha_{\rm D}}
\newcommand{\skin}{r_{\rm skin}}
\newcommand{\SV}{SV-min}
\newcommand{\SVb}{SV-min$^*$}

\newcommand{\RMF}{RMF-PC}
\newcommand{\RMFb}{RMF-PC$^*$}

\newcommand{\Pb}{$^{208}$Pb}
\newcommand{\Ca}{$^{48}$Ca}

\begin{document}

\title{Information content of the  parity-violating asymmetry in $^{208}$Pb}

\author{Paul-Gerhard Reinhard}
\affiliation{Institut für Theoretische Physik, Universität Erlangen, Erlangen, Germany}

\author{Xavier Roca-Maza}
\affiliation{Dipartimento di Fisica ``Aldo Pontremoli'', Universit\`a degli Studi di Milano, 20133 Milano, Italy and INFN, Sezione di Milano, 20133 Milano, Italy}

\author{Witold Nazarewicz}
\affiliation{Facility for Rare Isotope Beams and Department of Physics and Astronomy, Michigan State University, East Lansing, Michigan 48824, USA}

\date{\today}
\begin{abstract}
The parity violating asymmetry $\Apv$ in {\Pb}, recently measured by the PREX-2 collaboration, is studied using modern relativistic (covariant)  and non-relativistic energy density functionals. We first assess the theoretical uncertainty on $\Apv$ which is intrinsic to the adopted approach. 
To this end, we
use quantified  functionals that are able to accommodate our previous knowledge on  nuclear observables such as binding energies, charge radii, and the dipole polarizability $\aD$ of {\Pb}. 
We then add the quantified value of $\Apv$ together with $\aD$ to our  calibration dataset to optimize new functionals.
Based on these results, we predict a neutron skin thickness in {\Pb} $\skin =0.19\pm 0.02$\,fm
and the symmetry-energy slope  $L=54\pm 8$\,MeV. These values are consistent with other estimates based on astrophysical data and are significantly lower than those recently reported using a particular set of relativistic energy density functionals.  
We also make a prediction for the $\Apv$ value in {\Ca} that will be soon available from the CREX measurement.
\end{abstract}
\maketitle

{\it Introduction}.--- The recent measurement of the  parity-violating asymmetry $\Apv$ at transferred momentum $q=0.3978$/fm in {\Pb} by the PREX-2 collaboration \cite{PREX-2}
provided a highly anticipated observable that can inform models of nuclei and nuclear matter. In a separate theoretical paper \cite{Reed2021}, implications of the PREX-2 result on nuclear properties and the equation of state of neutron-rich matter have been discussed within a specific class of relativistic energy density functionals (EDFs). The authors relate the measured $\Apv$ to  $\skin$ and deduce from that  a rather large symmetry-energy slope parameter $L = 106\pm 37$ MeV and a large neutron skin thickness in {\Pb} $0.21\lesssim \skin\lesssim 0.31$ fm. The mean values of these quantities systematically overestimate the currently accepted limits \cite{Oertel2017,Roca-Maza2018,Li2021}. 

We emphasize the fact that the new experimental information provided by PREX-2 collaboration is the $\Apv$ measured at a specific kinematic condition. Other nuclear quantities of interest reported in \cite{PREX-2,Reed2021}, such as the neutral weak form-factor, neutron skin  thickness, interior weak density, interior baryon density, and symmetry energy parameters,  become accessible only via theoretical models.  

The question addressed in this Letter is whether the PREX-2 value of $\Apv$ creates a principle tension with other data and models, as claimed in \cite{Reed2021}. The strategy is, first, to study $\Apv$ directly rather than non-observable quantities, and second, to employ a broad set of structurally different EDFs together with a statistical analysis \cite{Dob14a} to estimate the uncertainty on $\Apv$ intrinsic to each EDF as well as the correlation with other observables. In particular, we consider the relation with the electric dipole polarizability $\aD$ in {\Pb} which is known to be strongly correlated with $\skin$  and weak form factor \cite{Reinhard2013,Nazarewicz2014,Erler2015} and for which independent experimental data exist \cite{Tamii2011,Roca-Maza2015}. All  EDFs under consideration show a clear correlation between $\Apv$ and $\aD$ and indicate a possible incompatibility of their current values. We extend the analysis to other observables as neutron skins, bulk symmetry energy and its slope, and we make predictions for $\Apv$ in {\Ca} at the CREX kinematics \cite{CREX}.

{\it The  parity-violating asymmetry.}---
$\Apv$ can be  obtained experimentally from longitudinally polarized elastic electron scattering \cite{Horowitz2001}. 
\begin{equation}
\Apv(Q^2)
 = \frac{d\sigma_R/d\Omega-d\sigma_L/d\Omega}{d\sigma_R/d\Omega+d\sigma_L/d\Omega},
\label{eq:APVex}
\end{equation}
where $d\sigma_L/d\Omega$  ($d\sigma_R/d\Omega$) is the differential cross section for the scattering of left (right) handed electrons, $\Omega$ is the solid angle, and $Q^2$ is the squared 
transferred four momentum. The scattering cross sections in (\ref{eq:APVex}), for a heavy nucleus, must be computed taking into account Coulomb distortions \cite{Horowitz1998,Roca-Maza2011}. To this end, we have modified the Dirac partial-wave code {\sc elsepa} \cite{Salvat2005} to deal with parity non-conserving potentials. Actually, the distribution of scattering angles in the PREX-2 experiment	has a non-negligible width which we  take into account by considering the PREX-2 acceptance function,  see	 supplemental material (SM)  \cite{SM} for details.

To gain insight into structure of the parity violating asymmetry, it is useful to inspect the Plane Wave Born Approximation expression for $\Apv$: \cite{Horowitz2001}
\begin{equation}
\Apv(Q^2)
  \approx
  \frac{G_FQ^2|Q^{(W)}_{N,Z}| }{4\sqrt{2}\,\pi\alpha Z}
  \;
  \frac{F_W(q)}{F_C(q)},
\label{eq:APVth}
\end{equation}
where $q=\sqrt{Q^2}$, $G_F=1.1663787\,10^{-5}/\mathrm{GeV}^2$ is the Fermi coupling constant, $F_W$ the weak form factor, $F_C$ is the charge form factor, and $Q^{(W)}_{N,Z}$ is the weak charge of the nucleus with $N$ neutrons and $Z$ protons. Since 
$F_C$ primarily depends on 
protons and  $F_W$ on neutrons, $\Apv$  decreases linearly with $\skin$ at low-$Q^2$, also when Coulomb distortions are taken into account \cite{Roca-Maza2011}. Consequently  this observable can be used to infer information on $\skin$.

Even if exploited at a single kinematic condition, $\Apv$ is one of the most promising observables to probe neutrons in nuclei since it is based on the well known electroweak interaction. Other promising observables (cf. Refs.~\cite{Reinhard2010,Roca-Maza2018,Roca-Maza2018_2}) sensitive to the neutron distribution in nuclei include the dipole polarizability $\aD$ \cite{Tamii2011,Roca-Maza2015}, which we shall discuss in this Letter.

{\it Error budget for $\Apv$}.--- In Table~\ref{tab:APV}, we list the  nucleonic parameters that are used for the calculation of the nucleon electromagnetic and weak form factors and $\Apv$, see SM \cite{SM} for details.

\begin{table}[htb]
\caption{\label{tab:APV}
Final choice of the parameters entering the calculation of
the weak form factor and $A_{PV}$:
the electric proton $\langle r_p^2\rangle$  and neutron $\langle r_n^2\rangle$  radii;  the  magnetic dipole moments, $\mu_p$ and $\mu_n$;  the strange quark electric coupling $\rho_s$  and the strange quark magnetic moment $\kappa_s$; the weak charge of neutrons $Q^{(W)}_n$ and protons $Q^{(W)}_p$; and the total weak charge of {\Pb} $Q_{126,82}^{(W)}$.
}
\begin{ruledtabular}
\begin{tabular}{lll}
 $\langle r_p^2\rangle$ (fm$^2$)
  & $0.726\pm 0.019$ & \cite{Atac2021}\\
 $\langle r_n^2\rangle$ (fm$^2$)
  & $-0.1161\pm 0.0022$ & \cite{PDG2018}\\
   $\mu_p$ & 2.792847  & \cite{PDG2018}\\
   $\mu_n$ & -1.9130  & \cite{PDG2018}\\
 $Q^{(W)}_p$ & $0.0713\pm 0.0001$ &\cite{Horowitz2012,Hoferichter2020} \\
 $Q^{(W)}_n$ & $-0.9888\pm 0.0011$ &\cite{Horowitz2012,Hoferichter2020} \\
 $\rho_s$ &  -0.24$\pm$0.70 & \cite{HAPPEX,Liu2007}\\
 $\kappa_s$ &  $-0.017\pm 0.004$ & \cite{Alexandrou2020}\\
 $Q_{126,82}^{(W)}$ & -117.9$\pm$0.3 & \cite{PREX-2,Gorchtein2020}\\
\end{tabular}
\end{ruledtabular}
\end{table}
   
\begin{figure}[!htb]
\includegraphics[width=1.0\columnwidth]{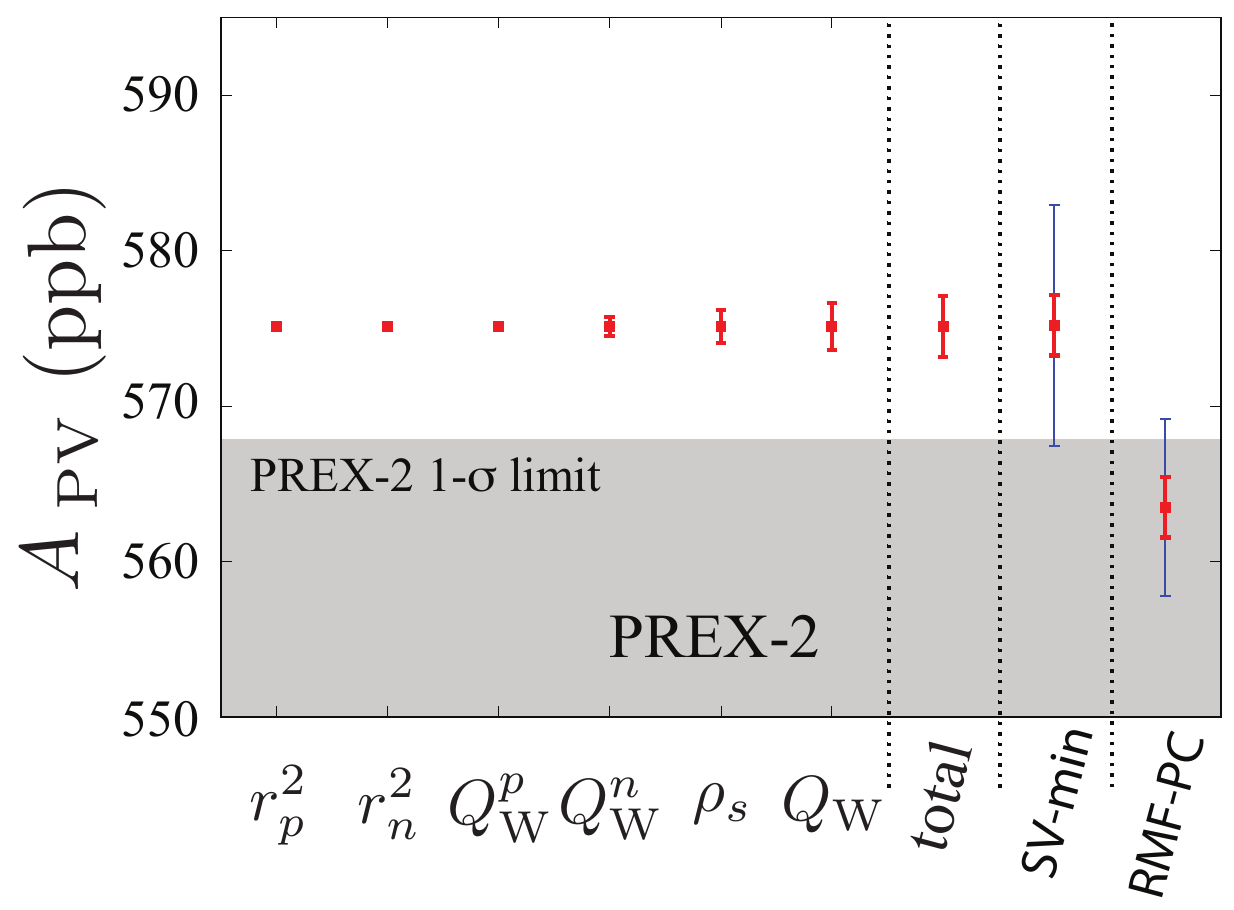}
\caption{\label{fig:APV_errors4} Uncertainty budget for  $\Apv$.  First six entries: the effect of the errors on the
  parameters in Table~\ref{tab:APV} on the uncertainty on $\Apv$.  The resulting total  uncertainty due to coupling constants
  is labeled ``total''. The quantified predictions  of $\Apv$ with SV-min and RMF-PC
  models (thin bars), which include statistical model uncertainties related to neutron and proton point densities and the coupling-constant uncertainty. The experimental value of $\Apv$ is 550$\pm$17.9\,ppb \cite{PREX-2}. The gray band marks the corresponding upper 1-sigma confidence interval.}
\end{figure}
Most parameters in Table~\ref{tab:APV} are given with errors either from experimental analysis or compilation of different sources. To estimate how these errors propagate to the prediction of $\Apv$ on a test calculation, we assume a  Gaussian profile  for the distribution of each parameter to sample the variance in $\Apv$. The result is shown in Fig.~\ref{fig:APV_errors4}. The first six entries show the impact of each parameter separately. Considerable contributions come only from the strength of the $s$ quark and, dominantly, from $Q_{N,Z}^{(W)}$. The entry ``total'' 
shows the total uncertainty from the first six entries accumulated by the Gaussian law of error propagation. 

There are also uncertainties on the predictions of the theoretical models (see below) stemming from the empirical calibration of the model parameters. The last two entries in Fig.~\ref{fig:APV_errors4} shows them (thin blue bars) for two typical model parametrizations discussed below together with the errors from the nucleonic parameters (thick red bars).  Both theoretical predictions are compatible, within errors, with the upper edge of the experimental uncertainty of the  PREX-2 measurement \cite{PREX-2}.

{\it Theoretical models} --
There exists a  variety of nuclear EDFs in the literature (for a
review see, e.g., \cite{Bender2003}). They differ in their 
structure and in the way there were calibrated.  We use here several families of EDFs having different functional form and
provide in similar fashion a set of parametrizations with systematically
varied symmetry energy $J$, while maintaining isoscalar properties and an overall good quality in their predictions.  This is of
particular interest when studying an observable like $\Apv$ which,
being related to the differences between the weak and electric charge densities, is predominantly sensitive
to the isovector channel of the EDFs \cite{Reinhard2013}.  The
families of EDFs considered in the survey are: FSU -- based on the traditional
non-linear Walecka model \cite{Serot1986} specially devised to
minimally improve its flexibility on the isovector channel
\cite{Piekarewicz2011}; RMF-DD and RMF-PC -- extended
relativistic mean-field models with more flexibility due to
density-dependent coupling constants.  DD employs the traditional
finite-range meson-exchange fields \cite{Niksic2002} while PC uses
point couplings \cite{Niksic2008}; the series of SV \cite{Klupfel2009}
and SAMi \cite{RocaMaza2012} parametrizations belong to the
widely used non-relativistic Skyrme EDFs; the
RD series is a variant of the Skyrme EDFs with a different form of
density dependence  \cite{Erler2010}.
Four of the families (SV, RD, PC, and DD) are calibrated to exactly
the same large set of ground observables: binding energies, charge radii,
diffraction radii, and surface thicknesses in semi-magic, spherical
nuclei \cite{Klupfel2009} plus a systematically
scanned constraint on symmetry energy $J$. The differences between the
results of these EDF families  show the impact of the EDF form.
The calibration is done by means of the standard linear regression, which also provides
information on  uncertainties and statistical correlations
between observables \cite{Dob14a,Erler2015}.  The other two families
(FSU and SAMi) are calibrated to different datasets with
different bias.  The SAMi functionals, e.g., have been optimized with the focus on  spin-isospin resonances.  We include these functionals to probe the impact of
calibration strategy. However, we checked that the performance for the reference nucleus, $^{208}$Pb, is roughly comparable for all parametrizations used, see the SM for details \cite{SM}. The inter-model comparison   helps quantifying the systematic theoretical error.

\begin{figure}[!htb]
\includegraphics[width=1.0\columnwidth]{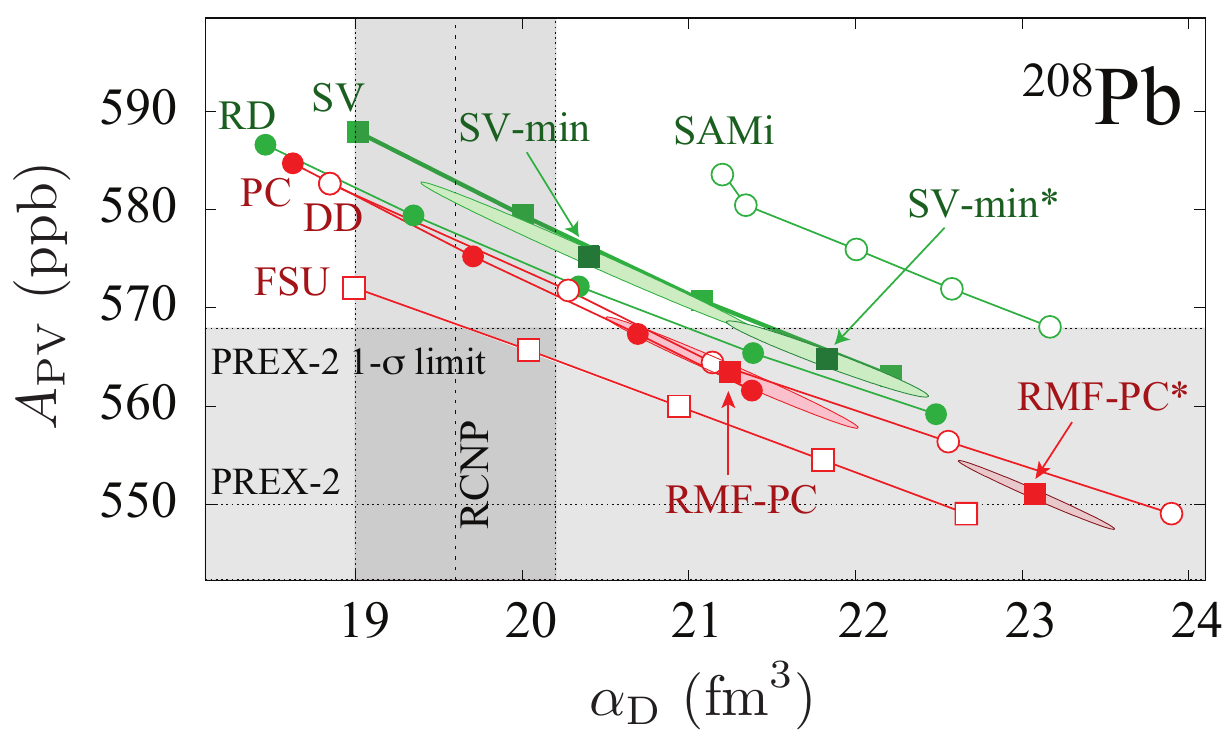}
\caption{\label{fig:APV-vs-polariz2} $\Apv$ versus
 $\aD$  in {\Pb}  for a set of covariant (red) and non-relativistic
 (green) EDFs.  Sets with systematically varied
  symmetry energy $J$ are connected by lines. (Note that $\aD$ increases as a function of $J$.)
  The {\SV}, {\SVb}, {\RMF}, and {\RMFb} results are shown together with their 1-sigma error ellipses. The experimental values of
  $\aD$ \cite{Tamii2011,Roca-Maza2015} and $\Apv$ \cite{PREX-2} are indicated together with their 1-sigma error bars.
}
\end{figure}
\begin{figure*}[!htb]
\includegraphics[width=0.9\linewidth]{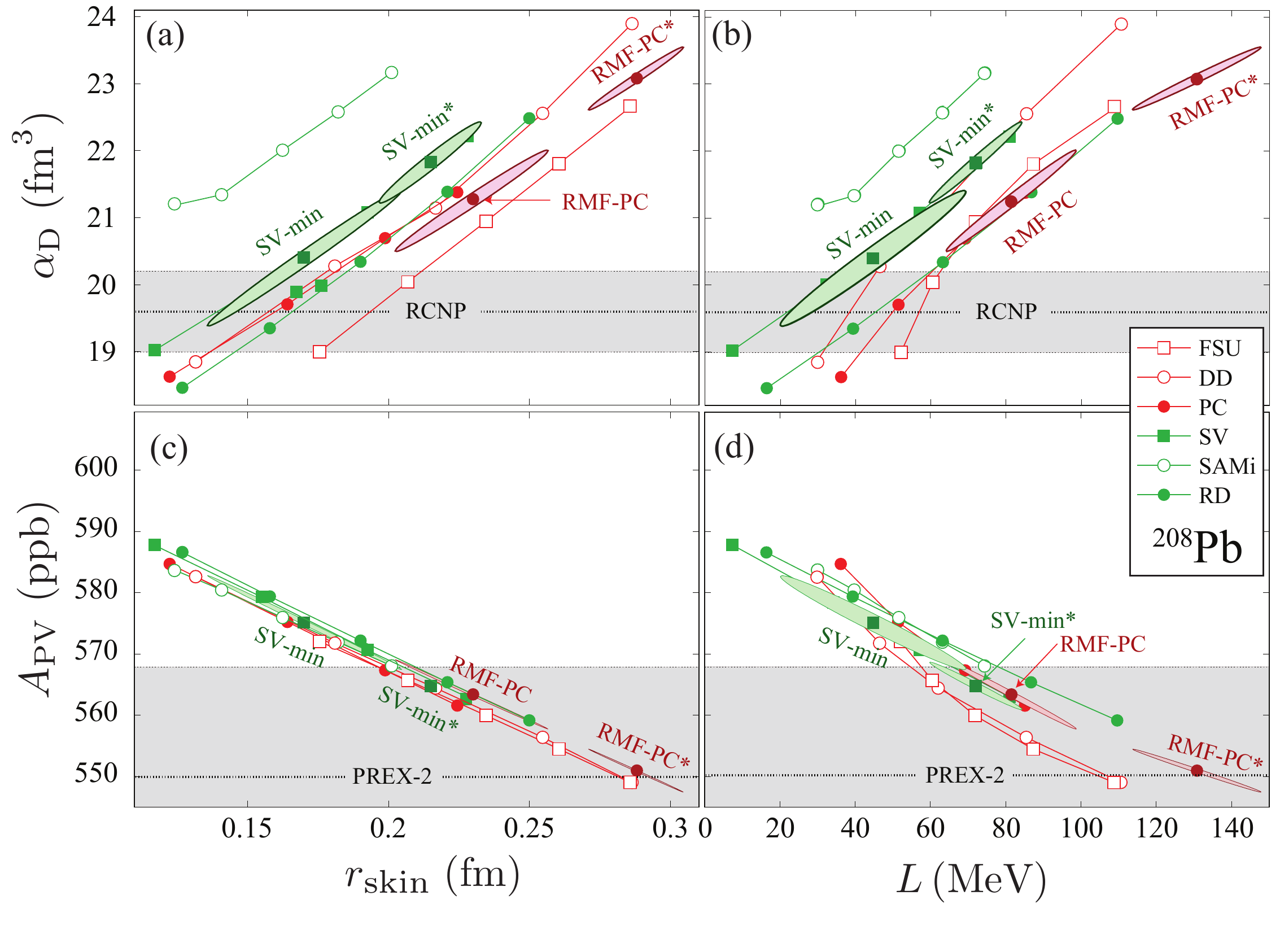}
\caption{\label{fig:APVaD} $\Apv$ (panels c,d)
  and   $\aD$  in {\Pb} (panels a,b)
  versus neutron skin (left panels) and slope of symmetry energy $L$ (right panels), for
  the same set of EDFs as in Fig.~\ref{fig:APV-vs-polariz2}. The experimental ranges of 
  $\aD$ \cite{Tamii2011,Roca-Maza2015} and $\Apv$ \cite{PREX-2} are marked.
  The values of $\skin$ (in fm) obtained in our models are: $0.17\pm 0.03$ for {\SV}; $0.22\pm 0.02$ for {\SVb}; $0.23\pm 0.03$ for {\RMF}; and $0.29\pm 0.02$  for {\RMFb}.
  The values of $L$ (in MeV) are: $45\pm 25$ for {\SV}; $72\pm 12$ for {\SVb}; $82\pm 17$ for {\RMF}; and $128\pm 17$ for {\RMFb}. 
}
\end{figure*}

{\it Tension between the PREX-2 result and electric dipole polarizability}.---The dipole polarizability $\aD$ in nuclei, directly related to the photo-absorption cross-section, provides an excellent constraint on $\skin$ \cite{Satula2006,Reinhard2010,Roca-Maza2013_2}.  The measurements of $\aD$ have been carried out for a number of nuclei, in particular for {\Pb} \cite{Tamii2011} and {\Ca} \cite{Birkhan2017}. 
These experiments provide a reliable information on the photo-absorption cross section up to about 20 MeV. Small high-energy contributions to $\aD$  require careful modeling of the quasi-deuteron  effect \cite{alphaD,schelhaas1988}, which motivated the correction from the original value $20.1 \pm 0.6$ fm$^3$ \cite{Tamii2011} to the value 19.6$\pm$0.6 fm$^3$ used here (cf. Ref.~\cite{Roca-Maza2015}). 

Figure~\ref{fig:APV-vs-polariz2} shows the predicted values of $\Apv$ versus $\aD$ obtained with the set of  covariant  and non-relativistic EDFs. The figure illustrates  a nearly linear trend of $\Apv$ versus $\aD$ with the same slope for all models, but slightly different offset mostly depending on  different values of the symmetry-energy coefficient $J$ predicted by the EDFs\,\cite{Roca-Maza2013_2}. The parametrizations {\SV} and {\RMF} stem from unconstrained fits to ground state data and their results are shown with the predicted 1-sigma error ellipses, which   align  along the average trend. This indicates that the statistical uncertainties of  {\SV} and {\RMF} are consistent with the systematic inter-model trends.
It is apparent there is only one model which is able to  reproduce simultaneously $\Apv$ and $\aD$ within the experimental 1-$\sigma$ error bands. The figure demonstrates therefore some tension: the models that are consistent with $\aD$ yield large values of $\Apv$ that are outside the 1-sigma limit of PREX-2 while the models that reproduce $\Apv$ yield  the values of $\aD$ that are well outside the experimental bounds.  The single model that seems to be  consistent with the current limits on $\Apv$ and  $\aD$  is the FSU EDF with  $J \sim 32$\,MeV  and  $L \sim 60$\,MeV, i.e., the symmetry energy that is well below the values advocated in Ref.~\cite{Reed2021}. Unfortunately, when it comes to other observables for {\Pb}, such as binding energy and charge radius, the performance of FSU models is inferior to the other EDFs discussed here, see \cite{SM} for details.

{\it New EDFs constrained on $\Apv$ and $\aD$}.---Figure~\ref{fig:APV-vs-polariz2} shows that the unconstrained fits, {\SV}  for the Skyrme functionals and {\RMF} for the RMF family, form a compromise between $\Apv$ and $\aD$ with the Skyrme functional tending  toward the mean value of $\aD$ and the RMF -- toward the mean value of $\Apv$.  To explore the compromise more systematically, we have fitted two new parametrizations taking the same set of ground state data from \cite{Klupfel2009} as were used for SV-min and PC-min and adding the experimental values for $\Apv$ and $\aD$ to the dataset of constraining observables. The relative weight of these two new data points is regulated by taking for the adopted errors the uncertainty of the model predictions from the unconstrained fits
(this amounts to 7\,ppb/5.7\,ppb for $\Apv$ and 1.0\,fm$^3$/0.7\,fm$^3$ for $\aD$ for SV-min/PC-min). 
We note that our adopted errors for $\Apv$ are close to the systematic error of PREX-2 measurement, which is 8\,ppb, and well below the statistical error of 16\,ppb.
The resulting parametrizations, called {\SVb} and {\RMFb}, stay on the general trend and move  toward the mean value of $\Apv$.
We also carried out optimizations assuming the  total experimental uncertainty of PREX-2  of 17.9\,ppb, dominated by statistics,   for the adopted error of $\Apv$. The   models calibrated under such assumption provide practically the same results as {\SV} and {\RMF} because the prior uncertainty on $\Apv$ is so large that the information content of this variable in this calibration scenario is low.
Based on Fig.~\ref{fig:APV-vs-polariz2} we conclude that  {\SVb} and {\RMFb} yield results that
are consistent with the current data on $\Apv$. On the other hand, the model  {\RMFb}, while closest to the mean value of  $\Apv$, is clearly inconsistent with the measured value of dipole polarizability.

{\it Symmetry energy and neutron skin}.---Over the years, strong correlations have been established between $\skin$ in heavy nuclei and various nuclear matter properties. Of particular importance, is the  correlation of $\skin$ with  the symmetry energy
at the saturation point $J$ \cite{Tondeur1984,Reinhard1999,Yoshida2004,Reinhard2010}  and with the slope of the bulk symmetry energy  $L$  \cite{Reinhard1999,Chen2005,Centelles2009}, see also Refs.~\cite{Brown2000,Typel2001,Furnstahl2002,Nazarewicz2014}. In addition to numerous inter-model comparisons published,
strong correlation between $L$, $J$,  and $\skin$ in medium-mass and heavy spherical closed-shell nuclei has been demonstrated by means of the statistical correlation analysis \cite{Reinhard2010,Kortelainen2013,Reinhard2016R}. 
One can conclude from the previous body of work that  the models with large symmetry energy parameters $J$ and $L$ predict
smaller $\Apv$ and large $\aD$, as indicated by the trend shown in  Fig.~\ref{fig:APV-vs-polariz2}. Also, the relativistic models tend to yield stiffer (larger value of $L$) neutron equation of state compared to the non-relativistic models \cite{Typel2001,Piekarewicz2012}.

Figure~\ref{fig:APVaD} shows the model predictions as functions of $\skin$ and $L$ for the models employed.
 Our result for $J$ can be found in SM \cite{SM}.
There is one more important aspect in Fig.~\ref{fig:APVaD}(c): the trend of $\Apv$ versus $\skin$ has the by far smallest spread within the families of the models employed. This intimate connection is also confirmed by statistical analysis for SV and RMF-PC EDFs: the correlation coefficient between $\Apv$ and $\skin$ is 99.9\%.

It is interesting to compare the  values of symmetry energy predicted in this work with the current estimates based on astrophysical constraints \cite{Oertel2017,Li2021,Biswas2021} and chiral effective field theory \cite{Drischler2020,Essick2021}. To this end, we go back to Fig.~\ref{fig:APV-vs-polariz2} and search for those parametrization in each series (SV, RD, PC, DD) which comes closest to the intercept of the RNCP and PREX-2 band. The resulting  inter-model average  is our prediction and the corresponding variance becomes our estimate for the systematic model error. For the symmetry energy, this procedure yields $J=32\pm 1$ MeV. This value is consistent with 
$J=  31.6 \pm 2.7$\,MeV  \cite{Oertel2017}, $31.7 \pm 1.1$\,MeV \cite{Drischler2020},
34$\pm$3\,MeV \cite{Essick2021}, and well below the value of $J=38.1\pm4.7$\,MeV of Ref.~\cite{Reed2021}. 

The symmetry-energy slope is determined with larger uncertainty:$L=54\pm 8$\,MeV. This value is comparable with
$L=57.7\pm 19$\,MeV \cite{Li2021}, $69\pm 16$\,MeV \cite{Biswas2021}, $L=59.8\pm 4.1$\,MeV \cite{Drischler2020}, and $58\pm 19$\,MeV \cite{Essick2021}. The analysis of \cite{Reed2021} using specific relativistic EDFs  yields a fairly large value of $L=106\pm 37$\, MeV.

The models compatible with the experimental $\alpha_D$ for {\Pb} predict $\skin$  in the range $0.13-0.19$\,fm \cite{Piekarewicz2012,Roca-Maza2013_2, Roca-Maza2015}, i.e., in the range of {\SV} values. Our expectation for $\skin$ from the present analysis is $0.19\pm 0.02$\,fm, i.e., a mean value significantly lower than the estimate $0.283 \pm 0.071$\,fm of  Ref.~\cite{PREX-2}.

{\it CREX measurement of  $\Apv$ in} $^{48}$Ca.---The CREX measurement  will soon provide 
the highly anticipated data on $\Apv$ in {\Ca} \cite{CREX}. In SM \cite{SM} we discuss our predictions at the kinematic point of CREX $Q^2=0.03$\,GeV$^2$. 
Considering our results for $^{208}$Pb, we
chose the value for $\Apv$($^{48}$Ca) close to the prediction of {\SV}  with a slight bias toward
{\SVb},  which amounts to  $2400\pm 60$\,ppb. We note that our predictions of $\aD(^{48}\mathrm{Ca})$ are in a slight conflict with the current experimental estimate \cite{Birkhan2017}.

{\it Summary and perspectives}.---For the quantified EDFs, there exists a  tension between $\Apv$ and $\aD$. 
The functionals  {\SV}, {\SVb}, and {\RMF}  offer a reasonable compromise between the data on $\Apv$ and $\aD$; they also perform well for other properties of {\Pb}. According to our analysis, the significant 1-sigma uncertainty of PREX-2 value of $\Apv$ makes it difficult to use this observable as a meaningful  constraint on  the isovector sector of current EDFs.  On the other hand, our estimated model uncertainty on $\Apv$, 6-7\,ppb is close to the estimated systematic error of PREX-2 of 8\,ppb. We recommend this value for the future calibration studies. In this respect, 
the anticipated precision measurements of  $\Apv$  and $\aD$  will be extremely useful for the calibration of nuclear models.

The mean values of the symmetry-energy parameters $J$ and $L$,  and $\skin$ in {\Pb} predicted in this work   are  significantly lower than the estimates of  Refs.~\cite{PREX-2,Reed2021}. Our numbers are consistent with much of the previous work and the recent astrophysical estimates.

{\it Acknowledgements}.---Useful discussions with R.J. Furnstahl, C.J. Horowitz, K.S. Kumar, and D.R.  Phillips are gratefully acknowledged. X.R.M. thanks J. Erler and M. Gorchtein for useful discussions on the weak charge of ${}^{208}$Pb. This material is based upon work supported by the U.S.\ Department of Energy, Office of Science, Office of Nuclear Physics under award numbers DE-SC0013365 and DE-SC0018083 (NUCLEI SciDAC-4 collaboration).

\bibliography{references}

\begin{thebibliography}{54}%
\makeatletter
\providecommand \@ifxundefined [1]{%
 \@ifx{#1\undefined}
}%
\providecommand \@ifnum [1]{%
 \ifnum #1\expandafter \@firstoftwo
 \else \expandafter \@secondoftwo
 \fi
}%
\providecommand \@ifx [1]{%
 \ifx #1\expandafter \@firstoftwo
 \else \expandafter \@secondoftwo
 \fi
}%
\providecommand \natexlab [1]{#1}%
\providecommand \enquote  [1]{``#1''}%
\providecommand \bibnamefont  [1]{#1}%
\providecommand \bibfnamefont [1]{#1}%
\providecommand \citenamefont [1]{#1}%
\providecommand \href@noop [0]{\@secondoftwo}%
\providecommand \href [0]{\begingroup \@sanitize@url \@href}%
\providecommand \@href[1]{\@@startlink{#1}\@@href}%
\providecommand \@@href[1]{\endgroup#1\@@endlink}%
\providecommand \@sanitize@url [0]{\catcode `\\12\catcode `\$12\catcode
  `\&12\catcode `\#12\catcode `\^12\catcode `\_12\catcode `\%12\relax}%
\providecommand \@@startlink[1]{}%
\providecommand \@@endlink[0]{}%
\providecommand \url  [0]{\begingroup\@sanitize@url \@url }%
\providecommand \@url [1]{\endgroup\@href {#1}{\urlprefix }}%
\providecommand \urlprefix  [0]{URL }%
\providecommand \Eprint [0]{\href }%
\providecommand \doibase [0]{http://dx.doi.org/}%
\providecommand \selectlanguage [0]{\@gobble}%
\providecommand \bibinfo  [0]{\@secondoftwo}%
\providecommand \bibfield  [0]{\@secondoftwo}%
\providecommand \translation [1]{[#1]}%
\providecommand \BibitemOpen [0]{}%
\providecommand \bibitemStop [0]{}%
\providecommand \bibitemNoStop [0]{.\EOS\space}%
\providecommand \EOS [0]{\spacefactor3000\relax}%
\providecommand \BibitemShut  [1]{\csname bibitem#1\endcsname}%
\let\auto@bib@innerbib\@empty
\bibitem [{\citenamefont {Adhikari}\ \emph {et~al.}(2021)\citenamefont
  {Adhikari} \emph {et~al.}}]{PREX-2}%
  \BibitemOpen
  \bibfield  {author} {\bibinfo {author} {\bibfnamefont {D.}~\bibnamefont
  {Adhikari}} \emph {et~al.} (\bibinfo {collaboration} {PREX Collaboration}),\
  }\bibfield  {title} {\enquote {\bibinfo {title} {Accurate determination of
  the neutron skin thickness of $^{208}\mathrm{Pb}$ through parity-violation in
  electron scattering},}\ }\href {\doibase 10.1103/PhysRevLett.126.172502}
  {\bibfield  {journal} {\bibinfo  {journal} {Phys. Rev. Lett.}\ }\textbf
  {\bibinfo {volume} {126}},\ \bibinfo {pages} {172502} (\bibinfo {year}
  {2021})}\BibitemShut {NoStop}%
\bibitem [{\citenamefont {Reed}\ \emph {et~al.}(2021)\citenamefont {Reed},
  \citenamefont {Fattoyev}, \citenamefont {Horowitz},\ and\ \citenamefont
  {Piekarewicz}}]{Reed2021}%
  \BibitemOpen
  \bibfield  {author} {\bibinfo {author} {\bibfnamefont {B.~T.}\ \bibnamefont
  {Reed}}, \bibinfo {author} {\bibfnamefont {F.~J.}\ \bibnamefont {Fattoyev}},
  \bibinfo {author} {\bibfnamefont {C.~J.}\ \bibnamefont {Horowitz}}, \ and\
  \bibinfo {author} {\bibfnamefont {J.}~\bibnamefont {Piekarewicz}},\
  }\bibfield  {title} {\enquote {\bibinfo {title} {Implications of {PREX-2} on
  the equation of state of neutron-rich matter},}\ }\href {\doibase
  10.1103/PhysRevLett.126.172503} {\bibfield  {journal} {\bibinfo  {journal}
  {Phys. Rev. Lett.}\ }\textbf {\bibinfo {volume} {126}},\ \bibinfo {pages}
  {172503} (\bibinfo {year} {2021})}\BibitemShut {NoStop}%
\bibitem [{\citenamefont {Oertel}\ \emph {et~al.}(2017)\citenamefont {Oertel},
  \citenamefont {Hempel}, \citenamefont {Kl\"ahn},\ and\ \citenamefont
  {Typel}}]{Oertel2017}%
  \BibitemOpen
  \bibfield  {author} {\bibinfo {author} {\bibfnamefont {M.}~\bibnamefont
  {Oertel}}, \bibinfo {author} {\bibfnamefont {M.}~\bibnamefont {Hempel}},
  \bibinfo {author} {\bibfnamefont {T.}~\bibnamefont {Kl\"ahn}}, \ and\
  \bibinfo {author} {\bibfnamefont {S.}~\bibnamefont {Typel}},\ }\bibfield
  {title} {\enquote {\bibinfo {title} {Equations of state for supernovae and
  compact stars},}\ }\href {\doibase 10.1103/RevModPhys.89.015007} {\bibfield
  {journal} {\bibinfo  {journal} {Rev. Mod. Phys.}\ }\textbf {\bibinfo {volume}
  {89}},\ \bibinfo {pages} {015007} (\bibinfo {year} {2017})}\BibitemShut
  {NoStop}%
\bibitem [{\citenamefont {Roca-Maza}\ and\ \citenamefont
  {Paar}(2018)}]{Roca-Maza2018}%
  \BibitemOpen
  \bibfield  {author} {\bibinfo {author} {\bibfnamefont {X.}~\bibnamefont
  {Roca-Maza}}\ and\ \bibinfo {author} {\bibfnamefont {N.}~\bibnamefont
  {Paar}},\ }\bibfield  {title} {\enquote {\bibinfo {title} {Nuclear equation
  of state from ground and collective excited state properties of nuclei},}\
  }\href {\doibase 10.1016/j.ppnp.2018.04.001} {\bibfield  {journal} {\bibinfo
  {journal} {Prog. Part. Nucl. Phys.}\ }\textbf {\bibinfo {volume} {101}},\
  \bibinfo {pages} {96--176} (\bibinfo {year} {2018})}\BibitemShut {NoStop}%
\bibitem [{\citenamefont {Li}\ \emph {et~al.}(2021)\citenamefont {Li},
  \citenamefont {Cai}, \citenamefont {Xie},\ and\ \citenamefont
  {Zhang}}]{Li2021}%
  \BibitemOpen
  \bibfield  {author} {\bibinfo {author} {\bibfnamefont {B.-A.}\ \bibnamefont
  {Li}}, \bibinfo {author} {\bibfnamefont {B.-J.}\ \bibnamefont {Cai}},
  \bibinfo {author} {\bibfnamefont {W.-J.}\ \bibnamefont {Xie}}, \ and\
  \bibinfo {author} {\bibfnamefont {N.-B.}\ \bibnamefont {Zhang}},\ }\href
  {\doibase 10.3390/universe7060182} {\enquote {\bibinfo {title} {Progress in
  constraining nuclear symmetry energy using neutron star observables since
  {GW170817}},}\ } (\bibinfo {year} {2021})\BibitemShut {NoStop}%
\bibitem [{\citenamefont {Dobaczewski}\ \emph {et~al.}(2014)\citenamefont
  {Dobaczewski}, \citenamefont {Nazarewicz},\ and\ \citenamefont
  {Reinhard}}]{Dob14a}%
  \BibitemOpen
  \bibfield  {author} {\bibinfo {author} {\bibfnamefont {J.}~\bibnamefont
  {Dobaczewski}}, \bibinfo {author} {\bibfnamefont {W.}~\bibnamefont
  {Nazarewicz}}, \ and\ \bibinfo {author} {\bibfnamefont {P.-G.}\ \bibnamefont
  {Reinhard}},\ }\bibfield  {title} {\enquote {\bibinfo {title} {Error
  estimates of theoretical models: a guide},}\ }\href {\doibase
  10.1088/0954-3899/41/7/074001} {\bibfield  {journal} {\bibinfo  {journal} {J.
  Phys. G}\ }\textbf {\bibinfo {volume} {41}},\ \bibinfo {pages} {074001}
  (\bibinfo {year} {2014})}\BibitemShut {NoStop}%
\bibitem [{\citenamefont {Reinhard}\ \emph {et~al.}(2013)\citenamefont
  {Reinhard}, \citenamefont {Piekarewicz}, \citenamefont {Nazarewicz},
  \citenamefont {Agrawal}, \citenamefont {Paar},\ and\ \citenamefont
  {Roca-Maza}}]{Reinhard2013}%
  \BibitemOpen
  \bibfield  {author} {\bibinfo {author} {\bibfnamefont {P.-G.}\ \bibnamefont
  {Reinhard}}, \bibinfo {author} {\bibfnamefont {J.}~\bibnamefont
  {Piekarewicz}}, \bibinfo {author} {\bibfnamefont {W.}~\bibnamefont
  {Nazarewicz}}, \bibinfo {author} {\bibfnamefont {B.~K.}\ \bibnamefont
  {Agrawal}}, \bibinfo {author} {\bibfnamefont {N.}~\bibnamefont {Paar}}, \
  and\ \bibinfo {author} {\bibfnamefont {X.}~\bibnamefont {Roca-Maza}},\
  }\bibfield  {title} {\enquote {\bibinfo {title} {Information content of the
  weak-charge form factor},}\ }\href {\doibase 10.1103/PhysRevC.88.034325}
  {\bibfield  {journal} {\bibinfo  {journal} {Phys. Rev. C}\ }\textbf {\bibinfo
  {volume} {88}},\ \bibinfo {pages} {034325} (\bibinfo {year}
  {2013})}\BibitemShut {NoStop}%
\bibitem [{\citenamefont {Nazarewicz}\ \emph {et~al.}(2014)\citenamefont
  {Nazarewicz}, \citenamefont {Reinhard}, \citenamefont {Satu{\l}a},\ and\
  \citenamefont {Vretenar}}]{Nazarewicz2014}%
  \BibitemOpen
  \bibfield  {author} {\bibinfo {author} {\bibfnamefont {W.}~\bibnamefont
  {Nazarewicz}}, \bibinfo {author} {\bibfnamefont {P.~G.}\ \bibnamefont
  {Reinhard}}, \bibinfo {author} {\bibfnamefont {W.}~\bibnamefont {Satu{\l}a}},
  \ and\ \bibinfo {author} {\bibfnamefont {D.}~\bibnamefont {Vretenar}},\
  }\bibfield  {title} {\enquote {\bibinfo {title} {Symmetry energy in nuclear
  density functional theory},}\ }\href {\doibase 10.1140/epja/i2014-14020-3}
  {\bibfield  {journal} {\bibinfo  {journal} {Eur. Phys. J. A}\ }\textbf
  {\bibinfo {volume} {50}},\ \bibinfo {pages} {20} (\bibinfo {year}
  {2014})}\BibitemShut {NoStop}%
\bibitem [{\citenamefont {Erler}\ and\ \citenamefont
  {Reinhard}(2015)}]{Erler2015}%
  \BibitemOpen
  \bibfield  {author} {\bibinfo {author} {\bibfnamefont {J.}~\bibnamefont
  {Erler}}\ and\ \bibinfo {author} {\bibfnamefont {P.-G.}\ \bibnamefont
  {Reinhard}},\ }\bibfield  {title} {\enquote {\bibinfo {title} {Error
  estimates for the {Skyrme-Hartree-Fock} model},}\ }\href {\doibase
  10.1088/0954-3899/42/3/034026} {\bibfield  {journal} {\bibinfo  {journal} {J.
  Phys. G}\ }\textbf {\bibinfo {volume} {42}},\ \bibinfo {pages} {034026}
  (\bibinfo {year} {2015})}\BibitemShut {NoStop}%
\bibitem [{\citenamefont {Tamii}\ \emph {et~al.}(2011)\citenamefont {Tamii}
  \emph {et~al.}}]{Tamii2011}%
  \BibitemOpen
  \bibfield  {author} {\bibinfo {author} {\bibfnamefont {A.}~\bibnamefont
  {Tamii}} \emph {et~al.},\ }\bibfield  {title} {\enquote {\bibinfo {title}
  {Complete electric dipole response and the neutron skin in
  $^{208}\mathrm{Pb}$},}\ }\href {\doibase 10.1103/PhysRevLett.107.062502}
  {\bibfield  {journal} {\bibinfo  {journal} {Phys. Rev. Lett.}\ }\textbf
  {\bibinfo {volume} {107}},\ \bibinfo {pages} {062502} (\bibinfo {year}
  {2011})}\BibitemShut {NoStop}%
\bibitem [{\citenamefont {Roca-Maza}\ \emph {et~al.}(2015)\citenamefont
  {Roca-Maza}, \citenamefont {Vi\~nas}, \citenamefont {Centelles},
  \citenamefont {Agrawal}, \citenamefont {Col\`o}, \citenamefont {Paar},
  \citenamefont {Piekarewicz},\ and\ \citenamefont {Vretenar}}]{Roca-Maza2015}%
  \BibitemOpen
  \bibfield  {author} {\bibinfo {author} {\bibfnamefont {X.}~\bibnamefont
  {Roca-Maza}}, \bibinfo {author} {\bibfnamefont {X.}~\bibnamefont {Vi\~nas}},
  \bibinfo {author} {\bibfnamefont {M.}~\bibnamefont {Centelles}}, \bibinfo
  {author} {\bibfnamefont {B.~K.}\ \bibnamefont {Agrawal}}, \bibinfo {author}
  {\bibfnamefont {G.}~\bibnamefont {Col\`o}}, \bibinfo {author} {\bibfnamefont
  {N.}~\bibnamefont {Paar}}, \bibinfo {author} {\bibfnamefont {J.}~\bibnamefont
  {Piekarewicz}}, \ and\ \bibinfo {author} {\bibfnamefont {D.}~\bibnamefont
  {Vretenar}},\ }\bibfield  {title} {\enquote {\bibinfo {title} {Neutron skin
  thickness from the measured electric dipole polarizability in
  $^{68}\text{Ni}$, $^{120}\text{Sn}$, and $^{208}\text{Pb}$},}\ }\href
  {\doibase 10.1103/PhysRevC.92.064304} {\bibfield  {journal} {\bibinfo
  {journal} {Phys. Rev. C}\ }\textbf {\bibinfo {volume} {92}},\ \bibinfo
  {pages} {064304} (\bibinfo {year} {2015})}\BibitemShut {NoStop}%
\bibitem [{\citenamefont {Riordan}\ \emph {et~al.}(2013)\citenamefont {Riordan}
  \emph {et~al.}}]{CREX}%
  \BibitemOpen
  \bibfield  {author} {\bibinfo {author} {\bibfnamefont {S.}~\bibnamefont
  {Riordan}} \emph {et~al.},\ }\bibfield  {title} {\enquote {\bibinfo {title}
  {{CREX: Parity violating measurement of the weak charge distribution of
  $^{48}${Ca} to 0.02 fm accuracy}},}\ }\href
  {http://www.jlab.org/exp_prog/proposals/13/C12-12-004_P.pdf} {\bibfield
  {journal} {\bibinfo  {journal} {JLAB-PR- 40-12-004, TJNAF, 2013}\ } (\bibinfo
  {year} {2013})}\BibitemShut {NoStop}%
\bibitem [{\citenamefont {Horowitz}\ \emph {et~al.}(2001)\citenamefont
  {Horowitz}, \citenamefont {Pollock}, \citenamefont {Souder},\ and\
  \citenamefont {Michaels}}]{Horowitz2001}%
  \BibitemOpen
  \bibfield  {author} {\bibinfo {author} {\bibfnamefont {C.~J.}\ \bibnamefont
  {Horowitz}}, \bibinfo {author} {\bibfnamefont {S.~J.}\ \bibnamefont
  {Pollock}}, \bibinfo {author} {\bibfnamefont {P.~A.}\ \bibnamefont {Souder}},
  \ and\ \bibinfo {author} {\bibfnamefont {R.}~\bibnamefont {Michaels}},\
  }\bibfield  {title} {\enquote {\bibinfo {title} {Parity violating
  measurements of neutron densities},}\ }\href {\doibase
  10.1103/PhysRevC.63.025501} {\bibfield  {journal} {\bibinfo  {journal} {Phys.
  Rev. C}\ }\textbf {\bibinfo {volume} {63}},\ \bibinfo {pages} {025501}
  (\bibinfo {year} {2001})}\BibitemShut {NoStop}%
\bibitem [{\citenamefont {Horowitz}(1998)}]{Horowitz1998}%
  \BibitemOpen
  \bibfield  {author} {\bibinfo {author} {\bibfnamefont {C.~J.}\ \bibnamefont
  {Horowitz}},\ }\bibfield  {title} {\enquote {\bibinfo {title} {Parity
  violating elastic electron scattering and {Coulomb} distortions},}\ }\href
  {\doibase 10.1103/PhysRevC.57.3430} {\bibfield  {journal} {\bibinfo
  {journal} {Phys. Rev. C}\ }\textbf {\bibinfo {volume} {57}},\ \bibinfo
  {pages} {3430--3436} (\bibinfo {year} {1998})}\BibitemShut {NoStop}%
\bibitem [{\citenamefont {Roca-Maza}\ \emph {et~al.}(2011)\citenamefont
  {Roca-Maza}, \citenamefont {Centelles}, \citenamefont {Vi\~nas},\ and\
  \citenamefont {Warda}}]{Roca-Maza2011}%
  \BibitemOpen
  \bibfield  {author} {\bibinfo {author} {\bibfnamefont {X.}~\bibnamefont
  {Roca-Maza}}, \bibinfo {author} {\bibfnamefont {M.}~\bibnamefont
  {Centelles}}, \bibinfo {author} {\bibfnamefont {X.}~\bibnamefont {Vi\~nas}},
  \ and\ \bibinfo {author} {\bibfnamefont {M.}~\bibnamefont {Warda}},\
  }\bibfield  {title} {\enquote {\bibinfo {title} {Neutron skin of
  $^{208}\mathrm{Pb}$, nuclear symmetry energy, and the parity radius
  experiment},}\ }\href {\doibase 10.1103/PhysRevLett.106.252501} {\bibfield
  {journal} {\bibinfo  {journal} {Phys. Rev. Lett.}\ }\textbf {\bibinfo
  {volume} {106}},\ \bibinfo {pages} {252501} (\bibinfo {year}
  {2011})}\BibitemShut {NoStop}%
\bibitem [{\citenamefont {Salvat}\ \emph {et~al.}(2005)\citenamefont {Salvat},
  \citenamefont {Jablonski},\ and\ \citenamefont {Powell}}]{Salvat2005}%
  \BibitemOpen
  \bibfield  {author} {\bibinfo {author} {\bibfnamefont {F.}~\bibnamefont
  {Salvat}}, \bibinfo {author} {\bibfnamefont {A.}~\bibnamefont {Jablonski}}, \
  and\ \bibinfo {author} {\bibfnamefont {C.~J.}\ \bibnamefont {Powell}},\
  }\bibfield  {title} {\enquote {\bibinfo {title} {{ELSEPA—Dirac}
  partial-wave calculation of elastic scattering of electrons and positrons by
  atoms, positive ions and molecules},}\ }\href {\doibase
  10.1016/j.cpc.2004.09.006} {\bibfield  {journal} {\bibinfo  {journal}
  {Comput. Phys. Comm.}\ }\textbf {\bibinfo {volume} {165}},\ \bibinfo {pages}
  {157--190} (\bibinfo {year} {2005})}\BibitemShut {NoStop}%
\bibitem [{SM()}]{SM}%
  \BibitemOpen
  \href@noop {} {}\bibinfo {note} {See Supplemental Material at [URL inserted
  by publisher] for more details on the angular averaging of $\Apv$, the weak
  charge of {\Pb}, electromagnetic and weak form factors, EDFs used in this
  work, and predictions for the symmetry energy coefficient $J$ in $^{208}$Pb
  and the electric dipole polarizability in $^{48}$Ca.}\BibitemShut {Stop}%
\bibitem [{\citenamefont {Reinhard}\ and\ \citenamefont
  {Nazarewicz}(2010)}]{Reinhard2010}%
  \BibitemOpen
  \bibfield  {author} {\bibinfo {author} {\bibfnamefont {P.-G.}\ \bibnamefont
  {Reinhard}}\ and\ \bibinfo {author} {\bibfnamefont {W.}~\bibnamefont
  {Nazarewicz}},\ }\bibfield  {title} {\enquote {\bibinfo {title} {Information
  content of a new observable: The case of the nuclear neutron skin},}\ }\href
  {\doibase 10.1103/PhysRevC.81.051303} {\bibfield  {journal} {\bibinfo
  {journal} {Phys. Rev. C}\ }\textbf {\bibinfo {volume} {81}},\ \bibinfo
  {pages} {051303} (\bibinfo {year} {2010})}\BibitemShut {NoStop}%
\bibitem [{\citenamefont {Roca-Maza}\ \emph {et~al.}(2018)\citenamefont
  {Roca-Maza}, \citenamefont {Col\`o},\ and\ \citenamefont
  {Sagawa}}]{Roca-Maza2018_2}%
  \BibitemOpen
  \bibfield  {author} {\bibinfo {author} {\bibfnamefont {X.}~\bibnamefont
  {Roca-Maza}}, \bibinfo {author} {\bibfnamefont {G.}~\bibnamefont {Col\`o}}, \
  and\ \bibinfo {author} {\bibfnamefont {H.}~\bibnamefont {Sagawa}},\
  }\bibfield  {title} {\enquote {\bibinfo {title} {Nuclear symmetry energy and
  the breaking of the isospin symmetry: How do they reconcile with each
  other?}}\ }\href {\doibase 10.1103/PhysRevLett.120.202501} {\bibfield
  {journal} {\bibinfo  {journal} {Phys. Rev. Lett.}\ }\textbf {\bibinfo
  {volume} {120}},\ \bibinfo {pages} {202501} (\bibinfo {year}
  {2018})}\BibitemShut {NoStop}%
\bibitem [{\citenamefont {Atac}\ \emph {et~al.}(2021)\citenamefont {Atac},
  \citenamefont {Constantinou}, \citenamefont {Meziani}, \citenamefont
  {Paolone},\ and\ \citenamefont {Sparveris}}]{Atac2021}%
  \BibitemOpen
  \bibfield  {author} {\bibinfo {author} {\bibfnamefont {H.}~\bibnamefont
  {Atac}}, \bibinfo {author} {\bibfnamefont {M.}~\bibnamefont {Constantinou}},
  \bibinfo {author} {\bibfnamefont {Z.~E.}\ \bibnamefont {Meziani}}, \bibinfo
  {author} {\bibfnamefont {M.}~\bibnamefont {Paolone}}, \ and\ \bibinfo
  {author} {\bibfnamefont {N.}~\bibnamefont {Sparveris}},\ }\bibfield  {title}
  {\enquote {\bibinfo {title} {Charge radii of the nucleon from its flavor
  dependent {Dirac} form factors},}\ }\href {\doibase
  10.1140/epja/s10050-021-00389-9} {\bibfield  {journal} {\bibinfo  {journal}
  {Eur. Phys. J. A}\ }\textbf {\bibinfo {volume} {57}},\ \bibinfo {pages} {65}
  (\bibinfo {year} {2021})}\BibitemShut {NoStop}%
\bibitem [{\citenamefont {Tanabashi}\ \emph {et~al.}(2018)\citenamefont
  {Tanabashi} \emph {et~al.}}]{PDG2018}%
  \BibitemOpen
  \bibfield  {author} {\bibinfo {author} {\bibfnamefont {M.}~\bibnamefont
  {Tanabashi}} \emph {et~al.} (\bibinfo {collaboration} {Particle Data
  Group}),\ }\bibfield  {title} {\enquote {\bibinfo {title} {Review of particle
  physics},}\ }\href {\doibase 10.1103/PhysRevD.98.030001} {\bibfield
  {journal} {\bibinfo  {journal} {Phys. Rev. D}\ }\textbf {\bibinfo {volume}
  {98}},\ \bibinfo {pages} {030001} (\bibinfo {year} {2018})}\BibitemShut
  {NoStop}%
\bibitem [{\citenamefont {Horowitz}\ and\ \citenamefont
  {Piekarewicz}(2012)}]{Horowitz2012}%
  \BibitemOpen
  \bibfield  {author} {\bibinfo {author} {\bibfnamefont {C.~J.}\ \bibnamefont
  {Horowitz}}\ and\ \bibinfo {author} {\bibfnamefont {J.}~\bibnamefont
  {Piekarewicz}},\ }\bibfield  {title} {\enquote {\bibinfo {title} {Impact of
  spin-orbit currents on the electroweak skin of neutron-rich nuclei},}\ }\href
  {\doibase 10.1103/PhysRevC.86.045503} {\bibfield  {journal} {\bibinfo
  {journal} {Phys. Rev. C}\ }\textbf {\bibinfo {volume} {86}},\ \bibinfo
  {pages} {045503} (\bibinfo {year} {2012})}\BibitemShut {NoStop}%
\bibitem [{\citenamefont {Hoferichter}\ \emph {et~al.}(2020)\citenamefont
  {Hoferichter}, \citenamefont {Men\'endez},\ and\ \citenamefont
  {Schwenk}}]{Hoferichter2020}%
  \BibitemOpen
  \bibfield  {author} {\bibinfo {author} {\bibfnamefont {M.}~\bibnamefont
  {Hoferichter}}, \bibinfo {author} {\bibfnamefont {J.}~\bibnamefont
  {Men\'endez}}, \ and\ \bibinfo {author} {\bibfnamefont {A.}~\bibnamefont
  {Schwenk}},\ }\bibfield  {title} {\enquote {\bibinfo {title} {Coherent
  elastic neutrino-nucleus scattering: {EFT} analysis and nuclear responses},}\
  }\href {\doibase 10.1103/PhysRevD.102.074018} {\bibfield  {journal} {\bibinfo
   {journal} {Phys. Rev. D}\ }\textbf {\bibinfo {volume} {102}},\ \bibinfo
  {pages} {074018} (\bibinfo {year} {2020})}\BibitemShut {NoStop}%
\bibitem [{\citenamefont {Acha}\ \emph {et~al.}(2007)\citenamefont {Acha} \emph
  {et~al.}}]{HAPPEX}%
  \BibitemOpen
  \bibfield  {author} {\bibinfo {author} {\bibfnamefont {A.}~\bibnamefont
  {Acha}} \emph {et~al.} (\bibinfo {collaboration} {HAPPEX Collaboration}),\
  }\bibfield  {title} {\enquote {\bibinfo {title} {Precision measurements of
  the nucleon strange form factors at ${Q}^{2}\sim 0.1$\, {GeV}$^{2}$},}\
  }\href {\doibase 10.1103/PhysRevLett.98.032301} {\bibfield  {journal}
  {\bibinfo  {journal} {Phys. Rev. Lett.}\ }\textbf {\bibinfo {volume} {98}},\
  \bibinfo {pages} {032301} (\bibinfo {year} {2007})}\BibitemShut {NoStop}%
\bibitem [{\citenamefont {Liu}\ \emph {et~al.}(2007)\citenamefont {Liu},
  \citenamefont {McKeown},\ and\ \citenamefont {Ramsey-Musolf}}]{Liu2007}%
  \BibitemOpen
  \bibfield  {author} {\bibinfo {author} {\bibfnamefont {J.}~\bibnamefont
  {Liu}}, \bibinfo {author} {\bibfnamefont {R.~D.}\ \bibnamefont {McKeown}}, \
  and\ \bibinfo {author} {\bibfnamefont {M.~J.}\ \bibnamefont
  {Ramsey-Musolf}},\ }\bibfield  {title} {\enquote {\bibinfo {title} {Global
  analysis of nucleon strange form factors at low ${Q}^{2}$},}\ }\href
  {\doibase 10.1103/PhysRevC.76.025202} {\bibfield  {journal} {\bibinfo
  {journal} {Phys. Rev. C}\ }\textbf {\bibinfo {volume} {76}},\ \bibinfo
  {pages} {025202} (\bibinfo {year} {2007})}\BibitemShut {NoStop}%
\bibitem [{\citenamefont {Alexandrou}\ \emph {et~al.}(2020)\citenamefont
  {Alexandrou}, \citenamefont {Bacchio}, \citenamefont {Constantinou},
  \citenamefont {Finkenrath}, \citenamefont {Hadjiyiannakou}, \citenamefont
  {Jansen},\ and\ \citenamefont {Koutsou}}]{Alexandrou2020}%
  \BibitemOpen
  \bibfield  {author} {\bibinfo {author} {\bibfnamefont {C.}~\bibnamefont
  {Alexandrou}}, \bibinfo {author} {\bibfnamefont {S.}~\bibnamefont {Bacchio}},
  \bibinfo {author} {\bibfnamefont {M.}~\bibnamefont {Constantinou}}, \bibinfo
  {author} {\bibfnamefont {J.}~\bibnamefont {Finkenrath}}, \bibinfo {author}
  {\bibfnamefont {K.}~\bibnamefont {Hadjiyiannakou}}, \bibinfo {author}
  {\bibfnamefont {K.}~\bibnamefont {Jansen}}, \ and\ \bibinfo {author}
  {\bibfnamefont {G.}~\bibnamefont {Koutsou}},\ }\bibfield  {title} {\enquote
  {\bibinfo {title} {Nucleon strange electromagnetic form factors},}\ }\href
  {\doibase 10.1103/PhysRevD.101.031501} {\bibfield  {journal} {\bibinfo
  {journal} {Phys. Rev. D}\ }\textbf {\bibinfo {volume} {101}},\ \bibinfo
  {pages} {031501} (\bibinfo {year} {2020})}\BibitemShut {NoStop}%
\bibitem [{\citenamefont {{J. Erler and M. Gorchtein, private
  communication}}(2021)}]{Gorchtein2020}%
  \BibitemOpen
  \bibfield  {author} {\bibinfo {author} {\bibnamefont {{J. Erler and M.
  Gorchtein, private communication}}},\ }\href@noop {} {} (\bibinfo {year}
  {2021})\BibitemShut {NoStop}%
\bibitem [{\citenamefont {Bender}\ \emph {et~al.}(2003)\citenamefont {Bender},
  \citenamefont {Heenen},\ and\ \citenamefont {Reinhard}}]{Bender2003}%
  \BibitemOpen
  \bibfield  {author} {\bibinfo {author} {\bibfnamefont {M.}~\bibnamefont
  {Bender}}, \bibinfo {author} {\bibfnamefont {P.-H.}\ \bibnamefont {Heenen}},
  \ and\ \bibinfo {author} {\bibfnamefont {P.-G.}\ \bibnamefont {Reinhard}},\
  }\bibfield  {title} {\enquote {\bibinfo {title} {Self-consistent mean-field
  models for nuclear structure},}\ }\href {\doibase 10.1103/RevModPhys.75.121}
  {\bibfield  {journal} {\bibinfo  {journal} {Rev. Mod. Phys.}\ }\textbf
  {\bibinfo {volume} {75}},\ \bibinfo {pages} {121--180} (\bibinfo {year}
  {2003})}\BibitemShut {NoStop}%
\bibitem [{\citenamefont {Serot}\ and\ \citenamefont
  {Walecka}(1986)}]{Serot1986}%
  \BibitemOpen
  \bibfield  {author} {\bibinfo {author} {\bibfnamefont {B.~D.}\ \bibnamefont
  {Serot}}\ and\ \bibinfo {author} {\bibfnamefont {J.~D.}\ \bibnamefont
  {Walecka}},\ }\bibfield  {title} {\enquote {\bibinfo {title} {The
  relativistic nuclear many--body problem},}\ }\href {\doibase
  10.1007/978-1-4684-5179-5_8} {\bibfield  {journal} {\bibinfo  {journal} {Adv.
  Nucl. Phys.}\ }\textbf {\bibinfo {volume} {16}},\ \bibinfo {pages} {1--316}
  (\bibinfo {year} {1986})}\BibitemShut {NoStop}%
\bibitem [{\citenamefont {Piekarewicz}(2011)}]{Piekarewicz2011}%
  \BibitemOpen
  \bibfield  {author} {\bibinfo {author} {\bibfnamefont {J.}~\bibnamefont
  {Piekarewicz}},\ }\bibfield  {title} {\enquote {\bibinfo {title} {Pygmy
  resonances and neutron skins},}\ }\href {\doibase 10.1103/PhysRevC.83.034319}
  {\bibfield  {journal} {\bibinfo  {journal} {Phys. Rev. C}\ }\textbf {\bibinfo
  {volume} {83}},\ \bibinfo {pages} {034319} (\bibinfo {year}
  {2011})}\BibitemShut {NoStop}%
\bibitem [{\citenamefont {Nik\v{s}i\'{c}}\ \emph {et~al.}(2002)\citenamefont
  {Nik\v{s}i\'{c}}, \citenamefont {Vretenar}, \citenamefont {Finelli},\ and\
  \citenamefont {Ring}}]{Niksic2002}%
  \BibitemOpen
  \bibfield  {author} {\bibinfo {author} {\bibfnamefont {T.}~\bibnamefont
  {Nik\v{s}i\'{c}}}, \bibinfo {author} {\bibfnamefont {D.}~\bibnamefont
  {Vretenar}}, \bibinfo {author} {\bibfnamefont {P.}~\bibnamefont {Finelli}}, \
  and\ \bibinfo {author} {\bibfnamefont {P.}~\bibnamefont {Ring}},\ }\bibfield
  {title} {\enquote {\bibinfo {title} {Relativistic {Hartree-Bogoliubov} model
  with density-dependent meson-nucleon couplings},}\ }\href {\doibase
  10.1103/PhysRevC.66.024306} {\bibfield  {journal} {\bibinfo  {journal} {Phys.
  Rev. C}\ }\textbf {\bibinfo {volume} {66}},\ \bibinfo {pages} {024306}
  (\bibinfo {year} {2002})}\BibitemShut {NoStop}%
\bibitem [{\citenamefont {Nik\v{s}i\'{c}}\ \emph {et~al.}(2008)\citenamefont
  {Nik\v{s}i\'{c}}, \citenamefont {Vretenar},\ and\ \citenamefont
  {Ring}}]{Niksic2008}%
  \BibitemOpen
  \bibfield  {author} {\bibinfo {author} {\bibfnamefont {T.}~\bibnamefont
  {Nik\v{s}i\'{c}}}, \bibinfo {author} {\bibfnamefont {D.}~\bibnamefont
  {Vretenar}}, \ and\ \bibinfo {author} {\bibfnamefont {P.}~\bibnamefont
  {Ring}},\ }\bibfield  {title} {\enquote {\bibinfo {title} {Relativistic
  nuclear energy density functionals: {Adjusting} parameters to binding
  energies},}\ }\href {\doibase 10.1103/PhysRevC.78.034318} {\bibfield
  {journal} {\bibinfo  {journal} {Phys. Rev. C}\ }\textbf {\bibinfo {volume}
  {78}},\ \bibinfo {pages} {034318} (\bibinfo {year} {2008})}\BibitemShut
  {NoStop}%
\bibitem [{\citenamefont {Kl\"upfel}\ \emph {et~al.}(2009)\citenamefont
  {Kl\"upfel}, \citenamefont {Reinhard}, \citenamefont {B\"urvenich},\ and\
  \citenamefont {Maruhn}}]{Klupfel2009}%
  \BibitemOpen
  \bibfield  {author} {\bibinfo {author} {\bibfnamefont {P.}~\bibnamefont
  {Kl\"upfel}}, \bibinfo {author} {\bibfnamefont {P.-G.}\ \bibnamefont
  {Reinhard}}, \bibinfo {author} {\bibfnamefont {T.~J.}\ \bibnamefont
  {B\"urvenich}}, \ and\ \bibinfo {author} {\bibfnamefont {J.~A.}\ \bibnamefont
  {Maruhn}},\ }\bibfield  {title} {\enquote {\bibinfo {title} {Variations on a
  theme by {Skyrme}: A systematic study of adjustments of model parameters},}\
  }\href {\doibase 10.1103/PhysRevC.79.034310} {\bibfield  {journal} {\bibinfo
  {journal} {Phys. Rev. C}\ }\textbf {\bibinfo {volume} {79}},\ \bibinfo
  {pages} {034310} (\bibinfo {year} {2009})}\BibitemShut {NoStop}%
\bibitem [{\citenamefont {Roca-Maza}\ \emph {et~al.}(2012)\citenamefont
  {Roca-Maza}, \citenamefont {Col\`o},\ and\ \citenamefont
  {Sagawa}}]{RocaMaza2012}%
  \BibitemOpen
  \bibfield  {author} {\bibinfo {author} {\bibfnamefont {X.}~\bibnamefont
  {Roca-Maza}}, \bibinfo {author} {\bibfnamefont {G.}~\bibnamefont {Col\`o}}, \
  and\ \bibinfo {author} {\bibfnamefont {H.}~\bibnamefont {Sagawa}},\
  }\bibfield  {title} {\enquote {\bibinfo {title} {New {Skyrme} interaction
  with improved spin-isospin properties},}\ }\href {\doibase
  10.1103/PhysRevC.86.031306} {\bibfield  {journal} {\bibinfo  {journal} {Phys.
  Rev. C}\ }\textbf {\bibinfo {volume} {86}},\ \bibinfo {pages} {031306}
  (\bibinfo {year} {2012})}\BibitemShut {NoStop}%
\bibitem [{\citenamefont {Erler}\ \emph {et~al.}(2010)\citenamefont {Erler},
  \citenamefont {Kl\"upfel},\ and\ \citenamefont {Reinhard}}]{Erler2010}%
  \BibitemOpen
  \bibfield  {author} {\bibinfo {author} {\bibfnamefont {J.}~\bibnamefont
  {Erler}}, \bibinfo {author} {\bibfnamefont {P.}~\bibnamefont {Kl\"upfel}}, \
  and\ \bibinfo {author} {\bibfnamefont {P.-G.}\ \bibnamefont {Reinhard}},\
  }\bibfield  {title} {\enquote {\bibinfo {title} {Exploration of a modified
  density dependence in the {Skyrme} functional},}\ }\href {\doibase
  10.1103/PhysRevC.82.044307} {\bibfield  {journal} {\bibinfo  {journal} {Phys.
  Rev. C}\ }\textbf {\bibinfo {volume} {82}},\ \bibinfo {pages} {044307}
  (\bibinfo {year} {2010})}\BibitemShut {NoStop}%
\bibitem [{\citenamefont {Satu\l{}a}\ \emph {et~al.}(2006)\citenamefont
  {Satu\l{}a}, \citenamefont {Wyss},\ and\ \citenamefont
  {Rafalski}}]{Satula2006}%
  \BibitemOpen
  \bibfield  {author} {\bibinfo {author} {\bibfnamefont {W.}~\bibnamefont
  {Satu\l{}a}}, \bibinfo {author} {\bibfnamefont {R.~A.}\ \bibnamefont {Wyss}},
  \ and\ \bibinfo {author} {\bibfnamefont {M.}~\bibnamefont {Rafalski}},\
  }\bibfield  {title} {\enquote {\bibinfo {title} {Global properties of the
  {Skyrme}-force-induced nuclear symmetry energy},}\ }\href {\doibase
  10.1103/PhysRevC.74.011301} {\bibfield  {journal} {\bibinfo  {journal} {Phys.
  Rev. C}\ }\textbf {\bibinfo {volume} {74}},\ \bibinfo {pages} {011301}
  (\bibinfo {year} {2006})}\BibitemShut {NoStop}%
\bibitem [{\citenamefont {Roca-Maza}\ \emph {et~al.}(2013)\citenamefont
  {Roca-Maza}, \citenamefont {Brenna}, \citenamefont {Col\`o}, \citenamefont
  {Centelles}, \citenamefont {Vi\~nas}, \citenamefont {Agrawal}, \citenamefont
  {Paar}, \citenamefont {Vretenar},\ and\ \citenamefont
  {Piekarewicz}}]{Roca-Maza2013_2}%
  \BibitemOpen
  \bibfield  {author} {\bibinfo {author} {\bibfnamefont {X.}~\bibnamefont
  {Roca-Maza}}, \bibinfo {author} {\bibfnamefont {M.}~\bibnamefont {Brenna}},
  \bibinfo {author} {\bibfnamefont {G.}~\bibnamefont {Col\`o}}, \bibinfo
  {author} {\bibfnamefont {M.}~\bibnamefont {Centelles}}, \bibinfo {author}
  {\bibfnamefont {X.}~\bibnamefont {Vi\~nas}}, \bibinfo {author} {\bibfnamefont
  {B.~K.}\ \bibnamefont {Agrawal}}, \bibinfo {author} {\bibfnamefont
  {N.}~\bibnamefont {Paar}}, \bibinfo {author} {\bibfnamefont {D.}~\bibnamefont
  {Vretenar}}, \ and\ \bibinfo {author} {\bibfnamefont {J.}~\bibnamefont
  {Piekarewicz}},\ }\bibfield  {title} {\enquote {\bibinfo {title} {Electric
  dipole polarizability in ${}^{208}${Pb}: {Insights} from the droplet
  model},}\ }\href {\doibase 10.1103/PhysRevC.88.024316} {\bibfield  {journal}
  {\bibinfo  {journal} {Phys. Rev. C}\ }\textbf {\bibinfo {volume} {88}},\
  \bibinfo {pages} {024316} (\bibinfo {year} {2013})}\BibitemShut {NoStop}%
\bibitem [{\citenamefont {Birkhan}\ \emph {et~al.}(2017)\citenamefont {Birkhan}
  \emph {et~al.}}]{Birkhan2017}%
  \BibitemOpen
  \bibfield  {author} {\bibinfo {author} {\bibfnamefont {J.}~\bibnamefont
  {Birkhan}} \emph {et~al.},\ }\bibfield  {title} {\enquote {\bibinfo {title}
  {Electric dipole polarizability of $^{48}\mathrm{Ca}$ and implications for
  the neutron skin},}\ }\href {\doibase 10.1103/PhysRevLett.118.252501}
  {\bibfield  {journal} {\bibinfo  {journal} {Phys. Rev. Lett.}\ }\textbf
  {\bibinfo {volume} {118}},\ \bibinfo {pages} {252501} (\bibinfo {year}
  {2017})}\BibitemShut {NoStop}%
\bibitem [{\citenamefont {Tamii}(2015)}]{alphaD}%
  \BibitemOpen
  \bibfield  {author} {\bibinfo {author} {\bibfnamefont {A.}~\bibnamefont
  {Tamii}},\ }\href@noop {} {} (\bibinfo {year} {2015}),\ \bibinfo {note}
  {private communication.}\BibitemShut {Stop}%
\bibitem [{\citenamefont {Schelhaas}\ \emph {et~al.}(1988)\citenamefont
  {Schelhaas}, \citenamefont {Henneberg}, \citenamefont {Sanzone-Arenhövel},
  \citenamefont {Wieloch-Laufenberg}, \citenamefont {Zurmühl}, \citenamefont
  {Ziegler}, \citenamefont {Schumacher},\ and\ \citenamefont
  {Wolf}}]{schelhaas1988}%
  \BibitemOpen
  \bibfield  {author} {\bibinfo {author} {\bibfnamefont {K.}~\bibnamefont
  {Schelhaas}}, \bibinfo {author} {\bibfnamefont {J.}~\bibnamefont
  {Henneberg}}, \bibinfo {author} {\bibfnamefont {M.}~\bibnamefont
  {Sanzone-Arenhövel}}, \bibinfo {author} {\bibfnamefont {N.}~\bibnamefont
  {Wieloch-Laufenberg}}, \bibinfo {author} {\bibfnamefont {U.}~\bibnamefont
  {Zurmühl}}, \bibinfo {author} {\bibfnamefont {B.}~\bibnamefont {Ziegler}},
  \bibinfo {author} {\bibfnamefont {M.}~\bibnamefont {Schumacher}}, \ and\
  \bibinfo {author} {\bibfnamefont {F.}~\bibnamefont {Wolf}},\ }\bibfield
  {title} {\enquote {\bibinfo {title} {Nuclear photon scattering by 208pb},}\
  }\href {\doibase https://doi.org/10.1016/0375-9474(88)90149-2} {\bibfield
  {journal} {\bibinfo  {journal} {Nuclear Physics A}\ }\textbf {\bibinfo
  {volume} {489}},\ \bibinfo {pages} {189--224} (\bibinfo {year}
  {1988})}\BibitemShut {NoStop}%
\bibitem [{\citenamefont {Tondeur}\ \emph {et~al.}(1984)\citenamefont
  {Tondeur}, \citenamefont {Brack}, \citenamefont {Farine},\ and\ \citenamefont
  {Pearson}}]{Tondeur1984}%
  \BibitemOpen
  \bibfield  {author} {\bibinfo {author} {\bibfnamefont {F.}~\bibnamefont
  {Tondeur}}, \bibinfo {author} {\bibfnamefont {M.}~\bibnamefont {Brack}},
  \bibinfo {author} {\bibfnamefont {M.}~\bibnamefont {Farine}}, \ and\ \bibinfo
  {author} {\bibfnamefont {J.}~\bibnamefont {Pearson}},\ }\bibfield  {title}
  {\enquote {\bibinfo {title} {Static nuclear properties and the
  parametrisation of {Skyrme} forces},}\ }\href {\doibase
  10.1016/0375-9474(84)90444-5} {\bibfield  {journal} {\bibinfo  {journal}
  {Nucl. Phys. A}\ }\textbf {\bibinfo {volume} {420}},\ \bibinfo {pages}
  {297--319} (\bibinfo {year} {1984})}\BibitemShut {NoStop}%
\bibitem [{\citenamefont {Reinhard}(1999)}]{Reinhard1999}%
  \BibitemOpen
  \bibfield  {author} {\bibinfo {author} {\bibfnamefont {P.-G.}\ \bibnamefont
  {Reinhard}},\ }\bibfield  {title} {\enquote {\bibinfo {title} {Skyrme forces
  and giant resonances in exotic nuclei},}\ }\href {\doibase
  10.1016/S0375-9474(99)00076-7} {\bibfield  {journal} {\bibinfo  {journal}
  {Nucl. Phys. A}\ }\textbf {\bibinfo {volume} {649}},\ \bibinfo {pages}
  {305--314} (\bibinfo {year} {1999})},\ \bibinfo {note} {giant
  Resonances}\BibitemShut {NoStop}%
\bibitem [{\citenamefont {Yoshida}\ and\ \citenamefont
  {Sagawa}(2004)}]{Yoshida2004}%
  \BibitemOpen
  \bibfield  {author} {\bibinfo {author} {\bibfnamefont {S.}~\bibnamefont
  {Yoshida}}\ and\ \bibinfo {author} {\bibfnamefont {H.}~\bibnamefont
  {Sagawa}},\ }\bibfield  {title} {\enquote {\bibinfo {title} {Neutron skin
  thickness and equation of state in asymmetric nuclear matter},}\ }\href
  {\doibase 10.1103/PhysRevC.69.024318} {\bibfield  {journal} {\bibinfo
  {journal} {Phys. Rev. C}\ }\textbf {\bibinfo {volume} {69}},\ \bibinfo
  {pages} {024318} (\bibinfo {year} {2004})}\BibitemShut {NoStop}%
\bibitem [{\citenamefont {Chen}\ \emph {et~al.}(2005)\citenamefont {Chen},
  \citenamefont {Ko},\ and\ \citenamefont {Li}}]{Chen2005}%
  \BibitemOpen
  \bibfield  {author} {\bibinfo {author} {\bibfnamefont {L.-W.}\ \bibnamefont
  {Chen}}, \bibinfo {author} {\bibfnamefont {C.~M.}\ \bibnamefont {Ko}}, \ and\
  \bibinfo {author} {\bibfnamefont {B.-A.}\ \bibnamefont {Li}},\ }\bibfield
  {title} {\enquote {\bibinfo {title} {Nuclear matter symmetry energy and the
  neutron skin thickness of heavy nuclei},}\ }\href {\doibase
  10.1103/PhysRevC.72.064309} {\bibfield  {journal} {\bibinfo  {journal} {Phys.
  Rev. C}\ }\textbf {\bibinfo {volume} {72}},\ \bibinfo {pages} {064309}
  (\bibinfo {year} {2005})}\BibitemShut {NoStop}%
\bibitem [{\citenamefont {Centelles}\ \emph {et~al.}(2009)\citenamefont
  {Centelles}, \citenamefont {Roca-Maza}, \citenamefont {Vi\~nas},\ and\
  \citenamefont {Warda}}]{Centelles2009}%
  \BibitemOpen
  \bibfield  {author} {\bibinfo {author} {\bibfnamefont {M.}~\bibnamefont
  {Centelles}}, \bibinfo {author} {\bibfnamefont {X.}~\bibnamefont
  {Roca-Maza}}, \bibinfo {author} {\bibfnamefont {X.}~\bibnamefont {Vi\~nas}},
  \ and\ \bibinfo {author} {\bibfnamefont {M.}~\bibnamefont {Warda}},\
  }\bibfield  {title} {\enquote {\bibinfo {title} {Nuclear symmetry energy
  probed by neutron skin thickness of nuclei},}\ }\href {\doibase
  10.1103/PhysRevLett.102.122502} {\bibfield  {journal} {\bibinfo  {journal}
  {Phys. Rev. Lett.}\ }\textbf {\bibinfo {volume} {102}},\ \bibinfo {pages}
  {122502} (\bibinfo {year} {2009})}\BibitemShut {NoStop}%
\bibitem [{\citenamefont {Alex~Brown}(2000)}]{Brown2000}%
  \BibitemOpen
  \bibfield  {author} {\bibinfo {author} {\bibfnamefont {B.}~\bibnamefont
  {Alex~Brown}},\ }\bibfield  {title} {\enquote {\bibinfo {title} {Neutron
  radii in nuclei and the neutron equation of state},}\ }\href {\doibase
  10.1103/PhysRevLett.85.5296} {\bibfield  {journal} {\bibinfo  {journal}
  {Phys. Rev. Lett.}\ }\textbf {\bibinfo {volume} {85}},\ \bibinfo {pages}
  {5296--5299} (\bibinfo {year} {2000})}\BibitemShut {NoStop}%
\bibitem [{\citenamefont {Typel}\ and\ \citenamefont
  {Brown}(2001)}]{Typel2001}%
  \BibitemOpen
  \bibfield  {author} {\bibinfo {author} {\bibfnamefont {S.}~\bibnamefont
  {Typel}}\ and\ \bibinfo {author} {\bibfnamefont {B.~A.}\ \bibnamefont
  {Brown}},\ }\bibfield  {title} {\enquote {\bibinfo {title} {Neutron radii and
  the neutron equation of state in relativistic models},}\ }\href {\doibase
  10.1103/PhysRevC.64.027302} {\bibfield  {journal} {\bibinfo  {journal} {Phys.
  Rev. C}\ }\textbf {\bibinfo {volume} {64}},\ \bibinfo {pages} {027302}
  (\bibinfo {year} {2001})}\BibitemShut {NoStop}%
\bibitem [{\citenamefont {Furnstahl}(2002)}]{Furnstahl2002}%
  \BibitemOpen
  \bibfield  {author} {\bibinfo {author} {\bibfnamefont {R.}~\bibnamefont
  {Furnstahl}},\ }\bibfield  {title} {\enquote {\bibinfo {title} {Neutron radii
  in mean-field models},}\ }\href {\doibase
  https://doi.org/10.1016/S0375-9474(02)00867-9} {\bibfield  {journal}
  {\bibinfo  {journal} {Nucl. Phys. A}\ }\textbf {\bibinfo {volume} {706}},\
  \bibinfo {pages} {85--110} (\bibinfo {year} {2002})}\BibitemShut {NoStop}%
\bibitem [{\citenamefont {Kortelainen}\ \emph {et~al.}(2013)\citenamefont
  {Kortelainen}, \citenamefont {Erler}, \citenamefont {Nazarewicz},
  \citenamefont {Birge}, \citenamefont {Gao},\ and\ \citenamefont
  {Olsen}}]{Kortelainen2013}%
  \BibitemOpen
  \bibfield  {author} {\bibinfo {author} {\bibfnamefont {M.}~\bibnamefont
  {Kortelainen}}, \bibinfo {author} {\bibfnamefont {J.}~\bibnamefont {Erler}},
  \bibinfo {author} {\bibfnamefont {W.}~\bibnamefont {Nazarewicz}}, \bibinfo
  {author} {\bibfnamefont {N.}~\bibnamefont {Birge}}, \bibinfo {author}
  {\bibfnamefont {Y.}~\bibnamefont {Gao}}, \ and\ \bibinfo {author}
  {\bibfnamefont {E.}~\bibnamefont {Olsen}},\ }\bibfield  {title} {\enquote
  {\bibinfo {title} {Neutron-skin uncertainties of {Skyrme} energy density
  functionals},}\ }\href {\doibase 10.1103/PhysRevC.88.031305} {\bibfield
  {journal} {\bibinfo  {journal} {Phys. Rev. C}\ }\textbf {\bibinfo {volume}
  {88}},\ \bibinfo {pages} {031305} (\bibinfo {year} {2013})}\BibitemShut
  {NoStop}%
\bibitem [{\citenamefont {Reinhard}\ and\ \citenamefont
  {Nazarewicz}(2016)}]{Reinhard2016R}%
  \BibitemOpen
  \bibfield  {author} {\bibinfo {author} {\bibfnamefont {P.-G.}\ \bibnamefont
  {Reinhard}}\ and\ \bibinfo {author} {\bibfnamefont {W.}~\bibnamefont
  {Nazarewicz}},\ }\bibfield  {title} {\enquote {\bibinfo {title} {Nuclear
  charge and neutron radii and nuclear matter: {Trend analysis in Skyrme}
  density-functional-theory approach},}\ }\href {\doibase
  10.1103/PhysRevC.93.051303} {\bibfield  {journal} {\bibinfo  {journal} {Phys.
  Rev. C}\ }\textbf {\bibinfo {volume} {93}},\ \bibinfo {pages} {051303}
  (\bibinfo {year} {2016})}\BibitemShut {NoStop}%
\bibitem [{\citenamefont {Piekarewicz}\ \emph {et~al.}(2012)\citenamefont
  {Piekarewicz}, \citenamefont {Agrawal}, \citenamefont {Col\`o}, \citenamefont
  {Nazarewicz}, \citenamefont {Paar}, \citenamefont {Reinhard}, \citenamefont
  {Roca-Maza},\ and\ \citenamefont {Vretenar}}]{Piekarewicz2012}%
  \BibitemOpen
  \bibfield  {author} {\bibinfo {author} {\bibfnamefont {J.}~\bibnamefont
  {Piekarewicz}}, \bibinfo {author} {\bibfnamefont {B.~K.}\ \bibnamefont
  {Agrawal}}, \bibinfo {author} {\bibfnamefont {G.}~\bibnamefont {Col\`o}},
  \bibinfo {author} {\bibfnamefont {W.}~\bibnamefont {Nazarewicz}}, \bibinfo
  {author} {\bibfnamefont {N.}~\bibnamefont {Paar}}, \bibinfo {author}
  {\bibfnamefont {P.-G.}\ \bibnamefont {Reinhard}}, \bibinfo {author}
  {\bibfnamefont {X.}~\bibnamefont {Roca-Maza}}, \ and\ \bibinfo {author}
  {\bibfnamefont {D.}~\bibnamefont {Vretenar}},\ }\bibfield  {title} {\enquote
  {\bibinfo {title} {Electric dipole polarizability and the neutron skin},}\
  }\href {\doibase 10.1103/PhysRevC.85.041302} {\bibfield  {journal} {\bibinfo
  {journal} {Phys. Rev. C}\ }\textbf {\bibinfo {volume} {85}},\ \bibinfo
  {pages} {041302} (\bibinfo {year} {2012})}\BibitemShut {NoStop}%
\bibitem [{\citenamefont {Biswas}(2021)}]{Biswas2021}%
  \BibitemOpen
  \bibfield  {author} {\bibinfo {author} {\bibfnamefont {B.}~\bibnamefont
  {Biswas}},\ }\href@noop {} {\enquote {\bibinfo {title} {Impact of {PREX-II,
  the revised mass measurement of PSRJ0740+6620, and possible NICER}
  observation on the dense matter equation of state},}\ } (\bibinfo {year}
  {2021}),\ \Eprint {http://arxiv.org/abs/2105.02886} {arXiv:2105.02886
  [astro-ph.HE]} \BibitemShut {NoStop}%
\bibitem [{\citenamefont {Drischler}\ \emph {et~al.}(2020)\citenamefont
  {Drischler}, \citenamefont {Furnstahl}, \citenamefont {Melendez},\ and\
  \citenamefont {Phillips}}]{Drischler2020}%
  \BibitemOpen
  \bibfield  {author} {\bibinfo {author} {\bibfnamefont {C.}~\bibnamefont
  {Drischler}}, \bibinfo {author} {\bibfnamefont {R.~J.}\ \bibnamefont
  {Furnstahl}}, \bibinfo {author} {\bibfnamefont {J.~A.}\ \bibnamefont
  {Melendez}}, \ and\ \bibinfo {author} {\bibfnamefont {D.~R.}\ \bibnamefont
  {Phillips}},\ }\bibfield  {title} {\enquote {\bibinfo {title} {How well do we
  know the neutron-matter equation of state at the densities inside neutron
  stars? {A Bayesian} approach with correlated uncertainties},}\ }\href
  {\doibase 10.1103/PhysRevLett.125.202702} {\bibfield  {journal} {\bibinfo
  {journal} {Phys. Rev. Lett.}\ }\textbf {\bibinfo {volume} {125}},\ \bibinfo
  {pages} {202702} (\bibinfo {year} {2020})}\BibitemShut {NoStop}%
\bibitem [{\citenamefont {Essick}\ \emph {et~al.}(2021)\citenamefont {Essick},
  \citenamefont {Tews}, \citenamefont {Landry},\ and\ \citenamefont
  {Schwenk}}]{Essick2021}%
  \BibitemOpen
  \bibfield  {author} {\bibinfo {author} {\bibfnamefont {R.}~\bibnamefont
  {Essick}}, \bibinfo {author} {\bibfnamefont {I.}~\bibnamefont {Tews}},
  \bibinfo {author} {\bibfnamefont {P.}~\bibnamefont {Landry}}, \ and\ \bibinfo
  {author} {\bibfnamefont {A.}~\bibnamefont {Schwenk}},\ }\href@noop {}
  {\enquote {\bibinfo {title} {Astrophysical constraints on the symmetry energy
  and the neutron skin of $^{208}${Pb} with minimal modeling assumptions},}\ }
  (\bibinfo {year} {2021}),\ \Eprint {http://arxiv.org/abs/2102.10074}
  {arXiv:2102.10074 [nucl-th]} \BibitemShut {NoStop}%
\end{thebibliography}%


%

\end{document}


\title{Supplemental Material for ``Information content of the  parity-violating asymmetry in $^{208}$Pb''}

\author{Paul-Gerhard Reinhard}
\affiliation{Institut für Theoretische Physik, Universität Erlangen, Erlangen, Germany}

\author{Xavier Roca-Maza}
\affiliation{Dipartimento di Fisica ``Aldo Pontremoli'', Universit\`a degli Studi di Milano, 20133 Milano, Italy and INFN, Sezione di Milano, 20133 Milano, Italy}

\author{Witold Nazarewicz}
\affiliation{Facility for Rare Isotope Beams and Department of Physics and Astronomy, Michigan State University, East Lansing, Michigan 48824, USA}

\begin{abstract}
This supplemental material contains supplemental discussions and 
three supplemental figures. Section \ref{sec:angaver} explains the averaging over the experimental distribution of scattering angles. Sections~\ref{qw} and \ref{ff} describe  the computation of weak charge, and charge and weak form factors, respectively. In Sec.~\ref{models} and Fig.~\ref{Modelquality} we assess the ability of various EDFs  to reproduce the basic data for {\Pb}. The predictions for the symmetry energy parameter $J$
are contained in Sec.~\ref{symmJ} and Fig.~\ref{Jsymm}. Finally, our results for the parity-violating asymmetry $\Apv$ in {\Ca} are contained in Sec.~\ref{CRX} and Fig.~\ref{aDCa}.

\end{abstract}

\maketitle

\setcounter{figure}{0}
\setcounter{table}{0}
\renewcommand{\thefigure}{S\arabic{figure}}
\renewcommand{\thetable}{S\arabic{table}}
\renewcommand{\theequation}{S\arabic{equation}}
\renewcommand{\theHfigure}{Extended Data Fig. \thefigure}
\renewcommand{\theHtable}{Extended Data Table \thetable}
\renewcommand{\thesection}{S.\Roman{section}}

\section{Angular-averaged $\Apv$}
\label{sec:angaver}

The parity-violating asymmetry $\Apv$ is a function of the transferred
momentum $q$, or equivalently four momentum $Q$. These are functions
of the incoming electron beam energy $E_\mathrm{el}$ and the
scattering angle $\theta$. To be precise, one should view $\Apv$ in
general as function of $E_\mathrm{el}$ and $\theta$. In a first round,
one discusses $\Apv$ at the experimental conditions of mean beam
energy and average angle, as was done in the formal presentation in
the paper.

The beam energy in the PREX-2 experiment is well defined while the
scattering angle $\theta$ has non-negligible width and data analysis
took that explicitly into account \cite{PREX-2}. Our theoretical
analysis follows the same procedure. Indeed,  what we discuss in this Letter is, in fact, the angular-averaged asymmetry:
\begin{equation}
  \Apv
  =
  \frac{\int d\theta\sin(\theta)\epsilon(\theta)\frac{d\sigma}{d\Omega}(\theta)\Apv(\theta)}
       {\int d\theta\sin(\theta)\epsilon(\theta)\frac{d\sigma}{d\Omega}(\theta)},
\label{eq:average}
\end{equation}
where $\epsilon(\theta)$ is the angular acceptance function as published in the supplemental material of \cite{PREX-2} and
$d\sigma/d\Omega$ is the differential cross section.
All quantities here are taken in the laboratory frame because
$\epsilon(\theta)$ is given in that frame. We replace the integral by
summation over the grid points as given in the experimental angle
distribution,  at each grid point angle and beam energy carry out transformation to
the center of mass (cm) frame, feed that to the DWBA code, and insert the resulting
$d\sigma/d\Omega$ and $\Apv(\theta)$ into
Eq. (\ref{eq:average}). Having summed that over all grid point yields
finally the angular averaged $\Apv$.

The simpler alternative is to take the average scattering angle
$\theta=4.69^o$ as given in \cite{PREX-2} and to calculate $\Apv$
at that one point. The difference between these two procedures amounts
to about 12 ppb. This is small as compared to the typical values for
$^{208}$Pb, namely $\Apv\approx 550-590$ ppb, but non-negligible at
the present level of discussion. Similar relations are found for
$^{48}$Ca discussed below where the shift is about 150 ppb out
of 2400 ppb. Thus we use angle-averaged $\Apv$ everywhere.

\section{Weak charge of {\Pb}}\label{qw}

Within the Standard Model, to lowest order, $Q_{N,Z}^{(W)}=-N+Z[1-4\sin(\theta_W
)]$ where $\theta_W$ is the weak-mixing angle. The scaling with neutron and proton numbers has been recently confirmed in atomic parity violation experiments in Ytterbium isotopes \cite{Antypas2019}.  However, radiative corrections to  $Q_{N,Z}^{(W)}$ need to be included for precise experiments. Within a 0.1\% accuracy, these corrections modify the previous expression as follows: $Q_{N,Z}^{(W)}=NQ_n^{(W)}+ZQ_p^{(W)}$ \cite{PDG2018}; which imply $Q_{126,82}^{(W)}=-118.8$. The latest theoretical estimate which includes also many-body effects in the radiative corrections is $-117.9\pm 0.3$ \cite{Gorchtein2020} and this value is used in our work. This value, properly scaled, is in agreement with the most accurate atomic parity violation measurement to date, that of the weak charge in ${}^{133}$Cs, which slightly deviates ($1.5\sigma$) from the Standard Model prediction \cite{Dzuba2012}. 
For the case of interest here, a reduction from $NQ_n^{(W)}+ZQ_p^{(W)}$ to $Q_{N,Z}^{(W)}$ by about 0.7\% implies a reduction on $\Apv$ of about 4 ppb. The same reduction of $\Apv$ would be produced by a change in the neutron rms radius of only 0.05 fm.     

\section{Computation of charge and weak form factors}\label{ff}

The parity-violating asymmetry $\Apv$ given in Eq. (2) 
depends on the
nuclear charge form factor $F_C(q)$ and weak form factor
$F_W(q)$ that depend on local proton and neutron density
distributions, $\rho_p$ and $\rho_n$, respectively. Accounting for
magnetic contributions requires also the spin-orbit current
($\nabla\mathbf{J}$ for SHF) \cite{Reinhard2021so} or the tensor current
($\rho_{T,p/n}$ for RMF) \cite{Horowitz2012}. The proton and neutron
densities are normalized in the usual way: $\int d^3r\rho_p=Z$ and
$\int d^3r\rho_n=N$.  We assume spherically symmetric systems, i.e.,
$\rho(\mathbf{r})=\rho(r)$ where $r=|\mathbf{r}|$.  In general, $F(q)$
and $\rho(r)$ are connected through the Fourier transformation
\cite{Fri82a}
\begin{subequations}
\begin{eqnarray}
  F(q)
  &=&
  \int d^3r\,e^{\mathrm{i}\mathrm{q}\cdot\mathrm{r}}\rho(r)
  =
  4\pi\int_0^\infty dr\,r^2\,j_0(qr)\rho(r),
\\
  \rho(r)
  &=&
  \int\!\!\frac{d^3q}{8\pi^3}e^{-\mathrm{i}\mathrm{q}\cdot\mathrm{r}}F(q)
  =
  \frac{1}{2\pi^2}\!\!\int_0^\infty\!\!\!\!\! dq\,q^2\,j_0(qr)F(q).
\end{eqnarray}
\end{subequations}
The transformation applies to any local density, for 
protons $\rho_p\longleftrightarrow F_p$,
neutrons $\rho_n\longleftrightarrow F_n$,
and the weak density $\rho_W\longleftrightarrow F_W$.

We prefer to formulate the weak distributions in terms of the
form factor because the necessary folding operations become much
simpler in the Fourier space.  Charge and weak form factors, both normalized to one, can be written as:
\begin{subequations}
\label{eq:formfweak}
\begin{eqnarray}
  F_C^{\mbox{}}(q)
  &=&
  \frac{e^{a_\mathrm{cm}q^2}}{Z}
  \sum_{t=p,n}\!\!\big(G_{E,t}(q)F_t(q)\!+\!G_{M,t}(q)F_{t}^{(ls)}(q)\big)
  ,
\\
  F_W^{\mbox{}}(q)
  &=&
  \frac{e^{a_\mathrm{cm}q^2}}{ZQ^{(W)}_p+NQ^{(W)}_n}
  \sum_{t=p,n}\!\!\big(G_{E,t}^{(W)}(q)F_t(q)\!+\!G_{M,t}^{(W)}(q)F_{t}^{(ls)}(q)\big)
  ,
\label{eq:FW}
\end{eqnarray}
\end{subequations}
where $a_\mathrm{cm}$ a parameter for the center-of-mass (c.m.)
correction, see Eq. (\ref{eq:cmcor}).  The charge form factor is
expressed in terms of $G_{E/M,p}$ and $G_{E/M,n}$, the intrinsic
proton and neutron electromagnetic form factors.  The weak
form factor calls the weak intrinsic nucleon form factors. They
are expressed in terms of the electromagnetic intrinsic form factors
weighted with the nucleonic weak charges as:
\begin{subequations}
\begin{eqnarray}
  G_{E,p}^{(W)}
  &=&
  Q^{(W)}_pG_{E,p}
  \!+\!
  Q^{(W)}_nG_{E,n}
  \!+\!
  Q^{(W)}_nG_{E,s}
  ,
\\
  G_{E,n}^{(W)}
  &=&
  Q^{(W)}_nG_{E,p}
  \!+\!
  Q^{(W)}_pG_{E,n}
  \!+\!
  Q^{(W)}_nG_{E,s}
  ,
 \\
  G_{M,p}^{(W)}
  &=&
  Q^{(W)}_pG_{M,p}
  \!+\!
  Q^{(W)}_nG_{M,n}
  \!+\!
  Q^{(W)}_nG_{M,s}
  ,
\\
  G_{M,n}^{(W)}
  &=&
  Q^{(W)}_nG_{M,p}
  \!+\!
  Q^{(W)}_pG_{M,n}
  \!+\!
  Q^{(W)}_nG_{M,s}
  ,
 \\
  G_{E,s}(q)
  &=&
  \rho_s\frac{\hbar^2q^2/(4c^2m_N^2)}{1+4.97\,\hbar^2q^2/(4c^2m_N^2)},
\label{eq:Gs}
 \\
  G_{M,s}(q)
  &=&
  \kappa_s\frac{\hbar^2}{(4c^2m_N^2)}
  ,
\end{eqnarray}
where $m_N$ is the average nucleon mass.  Note that the weak form
factor employs one more entry as compared to the electromagnetic
form factor, namely the strange-quark electromagnetic form factor
$G_{E/M,s}$. Its parameters together with nucleonic weak charges
and nucleon radii are given in Table I.

There is a great variety of publications on the parametrization of the
intrinsic electromagnetic nucleon form factors $G_{E/M,t}$, see,
e.g., \cite{Friar1975,Walther1986,Kel04a,Hoferichter2020}.  The data
point of interest here corresponds to low momentum $0.3978$/fm, which
is not very sensitive to the subtleties of the full form factors. The
leading parameters at low $q$ are the nucleonic radii and magnetic
moments, and we parametrize the nucleonic form factors on terms of
these parameters in a way which follows as close as possible the fully
fledged forms (tested in comparison to the full Mainz form factors
\cite{Simon1980,Walther1986} as reviewed, e.g., in \cite{Bender2003}).
This reads
\begin{eqnarray}
  G_{Ep}(q)
  &=&
  \frac{1}
       {1+\frac{1}{6}\langle r_{Ep}^2\rangle q^2}
  \,
  \sqrt{\frac{1}
       {1+\frac{\hbar^2}{(2m_Nc)^2}}q^2}
\\
  G_{En}(q)
  &=&
  G_{En}(q) 
  = 
  \frac{\langle r_{En}^2\rangle}{\langle r_{En}^2\rangle^\mathrm{(Mainz)}}
  \,
  G_{En}^\mathrm{(Mainz)}(q)
  ,
\\
  G_{M,p}(q)
  &=&
  -(1+2\mu_p)\frac{\hbar^2}{(2m_Nc)^2}G_M^\mathrm{(S)}(q)
  ,
\end{eqnarray}
\end{subequations}
where $G_{En}^\mathrm{(Mainz)}(q)$ stands for the Mainz
parametrization having
$\langle{r}_{En}^2\rangle^\mathrm{(Mainz)}=-0.117$ fm$^2$.  This is a
way to maintain some information on the $q$-dependence of $G_{En}(q)$
while having full control over the neutron radius.

Finally, a word about the c.m. correction. 
There are different ways to take the c.m. correction into account
in nuclear EDFs. Most functionals considered in the paper 
subtract the c.m. energy $\langle\hat{P}_{c.m.}^2\rangle/(2m_NA)$
a posteriori. In that case, the corresponding correction on radii
uses the factor
\begin{equation}
  a_\mathrm{cm}
  =
  \frac{\hbar^2}{8\langle\hat{P}_{c.m.}^2\rangle}
  .
\label{eq:cmcor}
\end{equation}
Some EDF use only the diagonal elements of $\hat{P}_{c.m.}^2$ which
allows to implement the c.m. correction by simple renormalization of
the nucleon mass $1/m_N\longrightarrow(1-1/A)/m_n$. In this case, the
effect on the form factor is already accounted for by the modified
kinetic energy and $a_\mathrm{cm}$ is set to zero.

\begin{figure}[!htb]
\includegraphics[width=0.6\columnwidth]{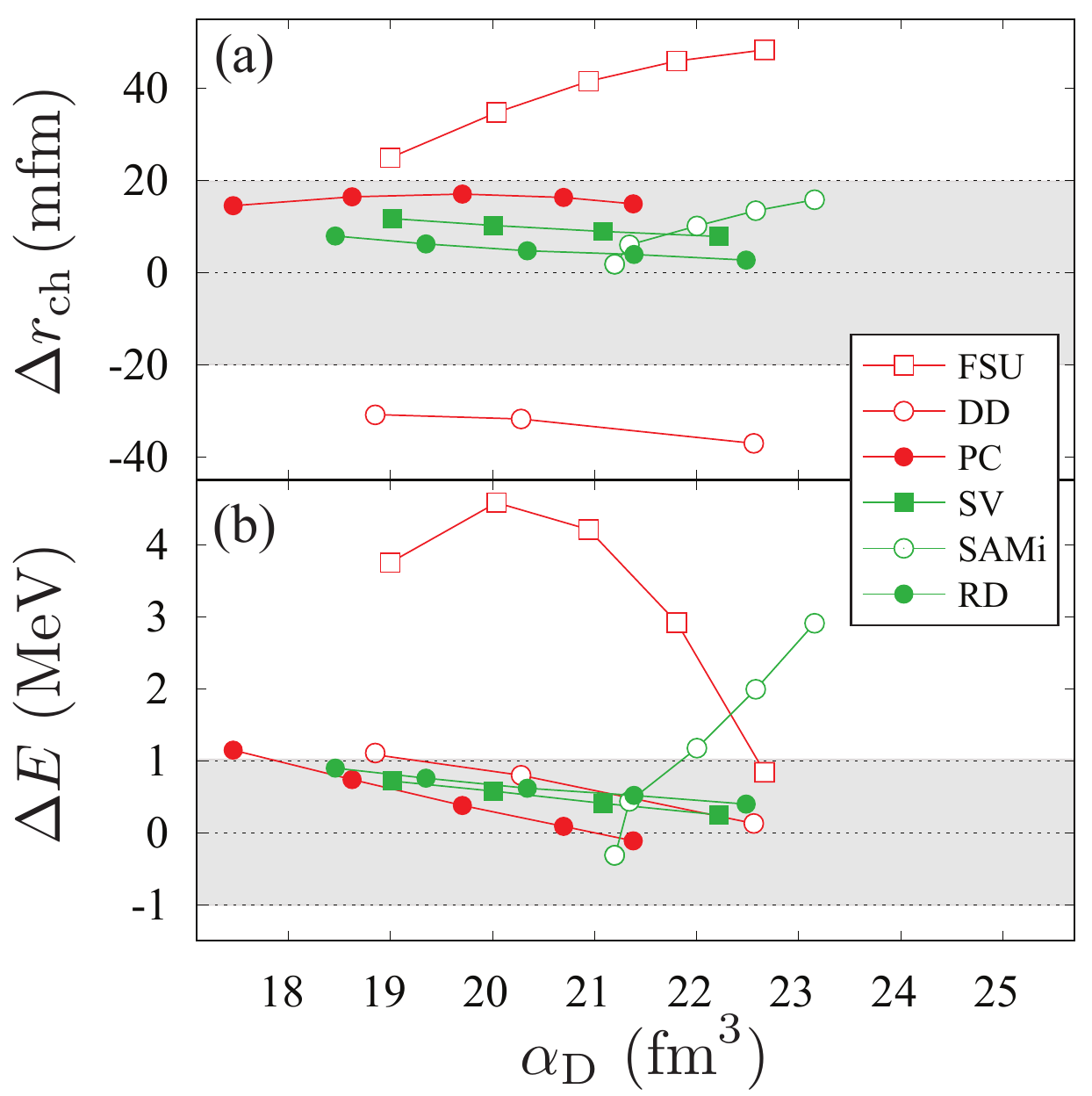}
\caption{The residuals of the charge radius (a)  and binding energy (b) of {\Pb} for the theoretical models used in this study. The grey bands around the perfect match indicate the typical performance of well adapted modern EDFs, i.e. the r.m.s. deviation taken over all nuclei where correlation effects are small \cite{Kluepfel2008,Klupfel2009}.}\label{Modelquality}
\end{figure}

\section{Ability of theoretical models to describe $^{208}$\text{Pb}}\label{models}

In this Letter, we compare the results for a  variety of EDFs. For
a fair comparison, these EDFs should also perform with approximately
similar quality for basic nuclear observables. As a minimal request,
we ask for comparable performance for the nucleus under consideration,
$^{208}$Pb. 

Figure~\ref{Modelquality} shows the deviation from experimental binding
energy (lower panel) and charge  radius (upper panel) for the
sets of parametrizations used in the paper. The grey error bands
indicate the typical r.m.s. error of up-to-date
parametrizations averaged over a broad selection of nuclei. 
The actual uncertainties indicated in Fig.~\ref{Modelquality}  are   $\sim$1\,MeV for the binding energy and $\sim$0.02\,fm for the charge radius.

The FSU family \cite{Piekarewicz2011}, still being
acceptable, falls out of the narrower range. This indicates the
limitations of the traditional non-linear RMF, even with the FSU
extensions. The RMF models with density dependent couplings (PC and DD
in the figure) were developed exactly for the purpose of allowing
better performance \cite{Niksic2002,Niksic2002}.

Models which fall outside the plot ranges were discarded for the
present survey. There are also other published EDFs which reproduce the {\Pb} data very well.
 We do not show them to render figures manageable.

\begin{figure}[!htb]
\includegraphics[width=0.6\columnwidth]{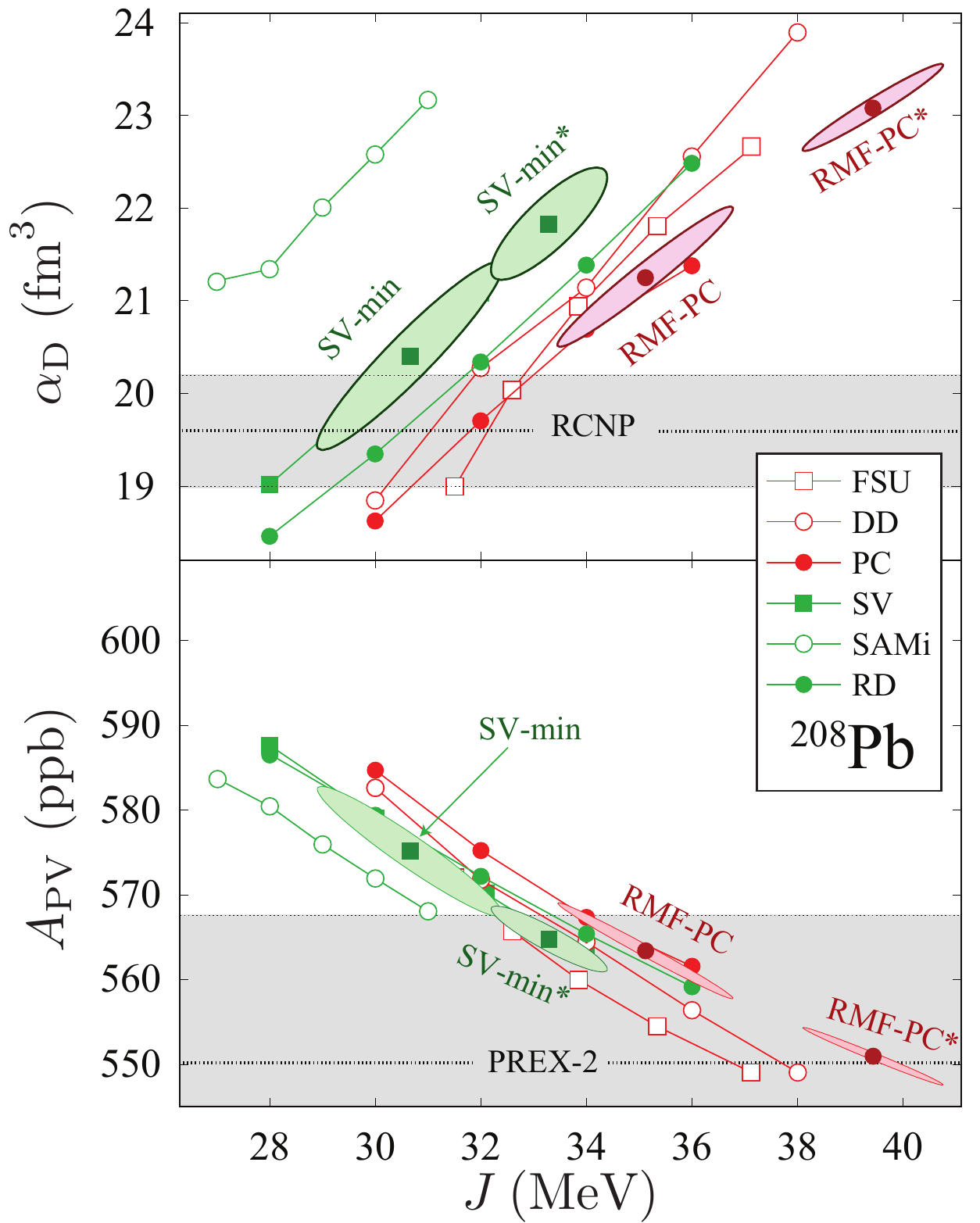}
\caption{Similar as in Fig. 3 but for the symmetry energy parameter $J$. The values of $J$ (in MeV) obtained in our models are: $31\pm 2$ for {\SV}; $33\pm 1$ for {\SVb}; $35\pm 2$ for {\RMF}; and $39\pm 1$ for {\RMFb}.
}\label{Jsymm}
\end{figure}

\section{Symmetry energy parameter $J$}\label{symmJ}

Figure~\ref{Jsymm} shows the predictions for $J$ by the  models employed.
Fig.~\ref{Jsymm} complements Fig.~3 of the paper by showing the trends
of $\Apv$ and $\aD$ with symmetry energy $J$. That looks very similar
to the trends for $L$; this   is not surprising as $J$ and $L$ are
very strongly correlated \cite{Reinhard2010,Reinhard2016R}.

\section{Parity-violating asymmetry in {\Ca}}\label{CRX}

\begin{figure}[!htb]
\includegraphics[width=0.6\columnwidth]{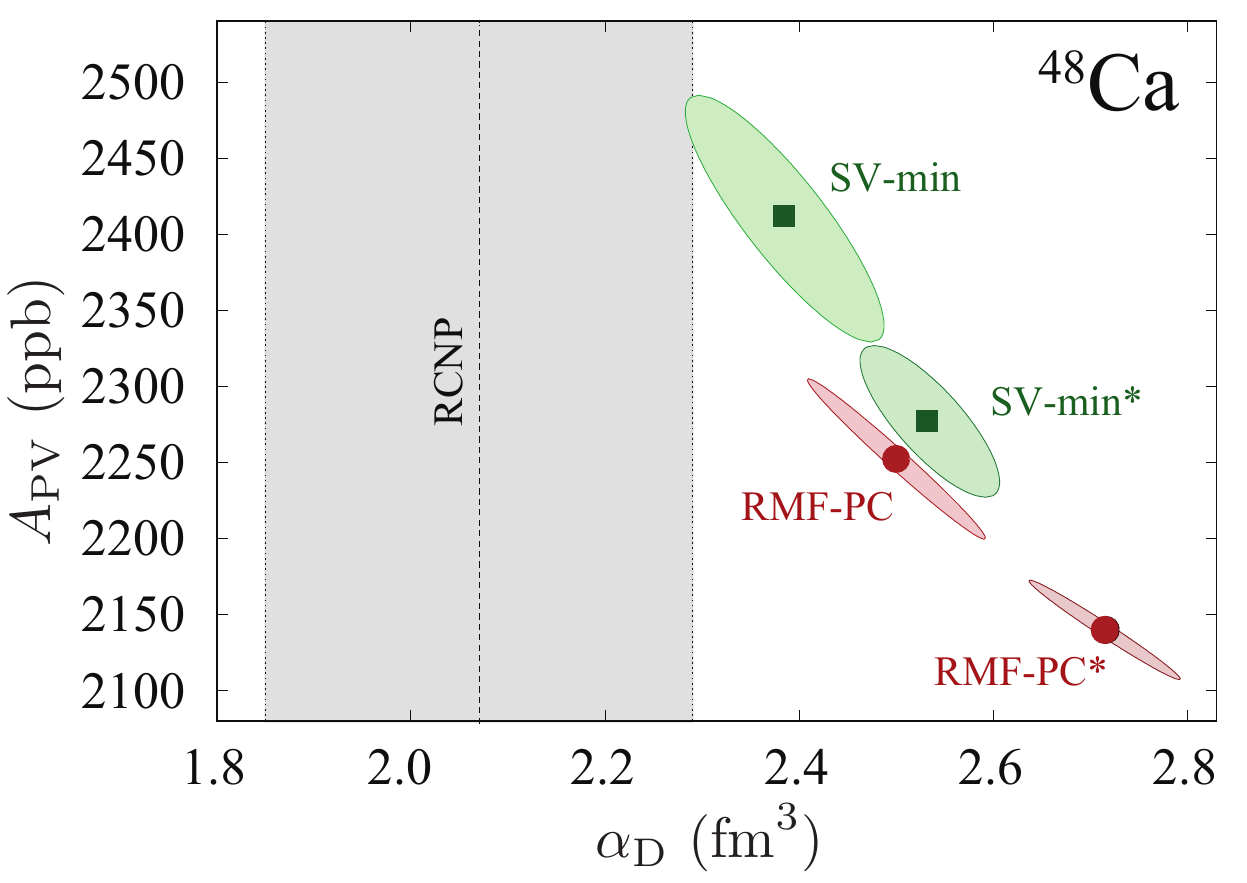}
\caption{$\Apv$ versus
 $\aD$  in {\Ca}  for  {\SV}, {\SVb}, {\RMF}, and {\RMFb}. The experimental value of
  $\aD$ \cite{Birkhan2017} is indicated.}\label{aDCa}
\end{figure}
With measurements of $\Apv$ in $^{48}$Ca being accomplished in near
future \cite{CREX,HorowitzC,Lin2015}, it is interesting to have a look at this
quantity. For composing $\Apv$, we use the same parameters as in Table I, except for the total weak charge $Q^{(W)}=-26.08$ which is deduced from $ZQ_p^{(W)}+NQ_n^{(W)}$ reduced by the factor 0.993 as in $^{208}$Pb. For averaging over scattering angles, we assume the same acceptance distribution as for the $^{208}$Pb experiment. These conditions may change in the final experiment which gives our prediction some principle uncertainty of a several dozen ppb.

Figure~\ref{aDCa} shows $\Apv$ versus $\aD$ in similar
fashion as for $^{208}$Pb in Fig.~2, however, restricted
to four parametrizations. 
 Note that the range of $\Apv$ shown here is
relatively narrower than in Fig.~2 for $^{208}$Pb. Based on
the prediction of {\SV},  with a slight bias toward
{\SVb}, we predict $\Apv(^{48}\mathrm{Ca})=2400\pm 60$\,ppb. This, again, may come in conflict with the current experimental data on
$\aD$ \cite{Birkhan2017}.

\bibliography{references}